\documentclass[aps,pra,reprint,superscriptaddress,longbibliography]{revtex4-1}
\pdfoutput=1
\usepackage{graphicx,color}
\usepackage{amsmath, amssymb}
\usepackage{longtable}
\usepackage{natbib}
\usepackage{bm}

\begin{document}
\title{
Zeeman interaction and Jahn-Teller effect in $\Gamma_8$ multiplet
}
\author{Naoya Iwahara}
\author{Veacheslav Vieru}
\author{Liviu Ungur}
\author{Liviu F. Chibotaru}
\email[]{Liviu.Chibotaru@chem.kuleuven.be}
\affiliation{Theory of Nanomaterials Group,
University of Leuven,
Celestijnenlaan 200F, B-3001 Leuven, Belgium}
\date{\today}

\begin{abstract}
We present a thorough analysis of the interplay of magnetic moment and the Jahn-Teller effect in the $\Gamma_8$ cubic multiplet. 
We find that in the presence of dynamical Jahn-Teller effect, the Zeeman interaction remains isotropic, whereas the $g$ and $G$ factors can change their signs. 
The static Jahn-Teller distortion also can change the sign of these $g$ factors as well as the nature of the magnetic anisotropy.
Combining the theory with state-of-the-art {\it ab initio} calculations, we analyzed the magnetic properties of Np$^{4+}$ and Ir$^{4+}$ impurity ions in cubic environment.
The calculated $g$ factors of Np$^{4+}$ impurity agree well with experimental data.
The {\it ab initio} calculation predicts strong Jahn-Teller effect in Ir$^{4+}$ ion in cubic environment and the strong vibronic reduction of $g$ and $G$ factors.
\end{abstract}


\maketitle

\section{Introduction}
\label{Sec:Introduction}
The magnetic complexes and insulators containing heavy transition metal, lanthanide, or actinide ions attract significant interest
because they often show single-molecular magnet behavior \cite{Sessoli2009, Layfield2015}
and various exotic magnetic phases \cite{Santini2009, Gingras2014, Witczak-Krempa2014}.
In these phenomena, local magnetic anisotropy with unquenched orbital angular momentum plays a key role.
Particularly, the magnetic interactions such as Zeeman interaction \cite{Abragam1970} and exchange interaction \cite{Elliott1968, Hartmann-Boutron1968}
become of tensorial type with the increase of the numbers of electronic states which are relevant to the magnetism.
At the same time, the vibronic coupling between the (quasi) degenerate electronic states and the lattice vibration \cite{Englman1972, Bersuker1989, Bersuker2006} leads to a more complex nature of the ground states.
Because of this, the analysis and the understanding of the electronic structure 
and magnetic properties of such materials is often incomplete.

The complexity already arises in local properties. 
The magnetic moment becomes nonlinear in pseudospin operators when at least four crystal field levels are involved \cite{Abragam1970, Chibotaru2012JCP, Chibotaru2013}.
Even in this minimal case corresponding to pseudospin $\tilde{S} = 3/2$ (the number of the crystal field levels $N = 2\tilde{S}+1$), 
each projection of magnetic moment is anisotropic and is described by as many as ten parameters. 
The number of parameters drastically decreases when the system has high spatial symmetry.
For example, in the case of cubic symmetry $(O, O_h, T_d)$, $\tilde{S}=3/2$ pseudospin state reduces to four-fold degenerate $\Gamma_8$ multiplet state, and the magnetic moment is expressed by only two parameters ($g$ and $G$) \cite{Abragam1970}.
Similarly, the number of vibronic coupling parameters reduces in cubic environment, while the lattice dynamics becomes of Jahn-Teller (JT) type \cite{Jahn1938, Englman1972, Bersuker1989}.
The $\Gamma_8$ state appears in many systems: single ion metal complexes \cite{Claassen1970, Osborne1978}, atomic clusters \cite{Sato2014, Domcke2016},
and impurities in insulators and semiconductors \cite{Abragam1970, Morgan1970, Bray1978, Bernstein1979, Edelstein1980, Halliday1988, Karbowiak1998}.
It also arises in magnetic sites of insulators such as double perovskites containing lanthanide or heavy transition metal ions 
\cite{Zhou2006, Erickson2007, deVries2010, Aharen2010, Marjerrison2016}, and lanthanide and actinide dioxides \cite{Webster2007, Santini2009}.
In order to get deep insight into the magnetic properties of these compounds, a thorough understanding of the complex behavior of the $\Gamma_8$ multiplet is of fundamental importance.

Even in cubic systems, it is usually not  easy to extract the parameters for Zeeman and vibronic interactions from experiment because of their complex interplay. 
On the other hand, {\it ab initio} methodology has proved to be a powerful tool for describing complex systems with localized electrons. 
With state-of-the-art quantum chemistry methodology, it is possible to obtain nowadays accurate low-energy crystal field states and, therefore, local magnetic moments and vibronic coupling constants \cite{Chap6}.
Besides being used in the field of molecular physics and theoretical chemistry, {\it ab initio} quantum chemistry methodology has been recently applied to the study of strongly correlated materials \cite{Bogdanov2015, Lafrancois2016}.
An important progress was the development of {\it ab initio} methodology for first principle calculations of anisotropic magnetic properties of single-ion metal centers and a unique derivation of pseudospin magnetic Hamiltonians for arbitrary multiplets \cite{Chibotaru2008, Chibotaru2012JCP, Chibotaru2013}.

In this work, we apply the {\it ab initio} approach to study the interplay of local vibronic and magnetic interactions in $\Gamma_8$ systems.
We find that in the presence of the JT dynamics, the Zeeman interaction remains isotropic, whereas the $g$ and $G$ factors describing the magnetic moments change their signs with respect to the pure electronic ones.
On the other hand, the static JT distortion not only makes the Zeeman interaction anisotropic in the two split Kramers doublets
but also changes the sign of $g$ factors in function of applied JT distortion. 
Finally, we calculate the Zeeman parameters for some realistic systems.

\section{Model Hamiltonian for $\Gamma_8$ multiplet}
\label{Sec:model}
Consider a metal ion with odd number of electrons in cubic environment, allowing for the existence of $\Gamma_8$ multiplets.
Hereafter, we assume that the components of $\Gamma_8$ multiplets $|\Phi_{\Gamma_8 M}\rangle$ ($M = -3/2, -1/2, 1/2, 3/2$) transform under symmetric operations as pure spin states $|S=3/2, M\rangle$ which are quantized along one of $C_4$ axes (the $z$ axis). 
According to the selection rules, 
\begin{eqnarray}
 [\Gamma_8^2] = a_2 \oplus 2t_1 \oplus t_2,
\label{Eq:selection1}
\end{eqnarray}
the $\Gamma_8$ multiplet has nonzero magnetic moment ($t_1$ representation in cubic groups) and is able, therefore, to respond linearly to applied magnetic field \cite{Abragam1970}. 
It also couples to vibrational modes whose symmetry is defined by the antisymmetric product of $\Gamma_8$ \cite{Jahn1938, Englman1972, Bersuker1989},
\begin{eqnarray}
 \{\Gamma_8^2\} = a_{1} \oplus e \oplus t_{2}.
\label{Eq:selection2}
\end{eqnarray}
Thus, the model Hamiltonian for the $\Gamma_8$ multiplet includes the Zeeman and the vibronic Hamiltonians. 

The magnetic moment is usually described by using pseudospin operators since the (spin, orbital, total) angular momenta do not commute with the Hamiltonian due to the concomitant presence of the crystal field and spin-orbit coupling \cite{Abragam1970}. 
The $\tilde{S}$ pseudospin states can be uniquely defined by unitary transformation of chosen $N$ crystal field states ($N = 2\tilde{S} + 1$). 
The imposed constraints are that the pseudospin states transform under time inversion and point group symmetry operations as true angular momentum states.
In the lack of sufficient symmetry and when $\tilde{S} > 1$, the pseudospin Hamiltonian can be defined via the adiabatic connection to well-established as true angular momentum states \cite{Chibotaru2008, Chibotaru2013}.
In the general case of $\tilde{S} = 3/2$ states without point group symmetry, the time-inversion symmetry reduces the number of the free parameters in the unitary transformation from 16 to 10 \cite{Chibotaru2013}.
Once the pseudospin states are defined, a time-odd operator in their space is expressed by linear combination of irreducible tensor operators, $Y_k^q(\tilde{\bm{S}})$ \cite{Varshalovich1988} of odd rank $k$, $k \le 2\tilde{S}$, where $\tilde{\bm{S}}$ is pseudospin operator. 
Accordingly, the magnetic moments $\mu_\alpha$ ($\alpha = x, y, z$) for $\tilde{S}=3/2$ states have the terms of rank $k = 1$ and $k = 3$, and they are expressed by 27 parameters \cite{Chibotaru2013}
\footnote{
Although the total number of independent parameter describing the three operators $\mu_\alpha$ is $10 \times 3 = 30$, in the case when the pseudospin $\tilde{S}$ is defined in a common Cartesian coordinate system, the $g$ tensor pasteurizing the first rank contribution to $\mu_\alpha$ become a symmetric matrix imposing three constraints on three parameters. 
}.
The latter reduce to 2 within the cubic symmetry when the four crystal field states correspond to $\Gamma_8$ multiplet: 
one component of the magnetic moments, say $\mu_z$, contains only $Y_1^0$ and $Y_3^0$ operators owing to the requirement of invariance under $C_4^z$ rotation,
thus, leaving only two parameters in its definition. 
Similarly, two parameters, $g_\alpha$ and $G_\alpha$, will define magnetic moments for other projections $\alpha$. 
Given the symmetry equivalence of tetragonal axes, we have $g_x = g_y = g_z = g$, and $G_x = G_y = G_z = G$. 
The existence of two independent parameters in the definition of $\bm{\mu}$ is also understood from the fact that $\Gamma_8$ is not simply reducible 
[the symmetric product $[\Gamma_8^2]$ contains two $t_1$ representations, Eq. (\ref{Eq:selection1})], hence, $t_1$ operators are expressed by two reduced matrix elements. 
Thus, the magnetic moment for the $\Gamma_8$ state is written as \cite{Bleaney1959, Abragam1970}:
\begin{eqnarray}
 \mu_\alpha &=&
  -\mu_\text{B} g O^\alpha_1(\tilde{\bm{S}}) - \mu_\text{B} G O_3^\alpha(\tilde{\bm{S}}),
\label{Eq:mu_Gamma8}
\end{eqnarray}
where, irreducible tensors $O_1^\alpha$, $O_3^\alpha$ are defined by
\begin{eqnarray}
 O_1^\alpha(\tilde{\bm{S}}) &=& \tilde{S}_\alpha,
\nonumber\\
 O_3^\alpha(\tilde{\bm{S}}) &=& \tilde{S}_\alpha^3 -\frac{1}{5} \tilde{S}_\alpha \left(3\tilde{S}(\tilde{S}+1)-1\right).
\label{Eq:O}
\end{eqnarray}

On the other hand, the linear Jahn-Teller Hamiltonian is \cite{Moffitt1957, Englman1972, Bersuker1989, Bersuker2006}:
\begin{eqnarray}
 H_{\rm JT} &=&
 \sum_{\Gamma \gamma}
 \left[
 \frac{1}{2}\left(P_{\Gamma\gamma}^2 + \omega_\Gamma^2 Q_{\Gamma\gamma}^2\right)I
 + V_\Gamma C_{\Gamma \gamma} Q_{\Gamma \gamma}
 \right],
\label{Eq:HJT}
\end{eqnarray}
where, $P_{E\gamma}$ $(\gamma = \theta, \epsilon)$ and $P_{T_2\gamma}$ $(\gamma = \xi, \eta, \zeta)$ are the conjugate momenta of
the mass-weighted normal coordinates $Q_{E\gamma}$ and $Q_{T_2\gamma}$, respectively, $\omega_\Gamma$ is the frequency of the $\Gamma$ mode ($\Gamma = E, T_2$),
$V_\Gamma$ are the linear vibronic coupling constants, $I$ is the $4 \times 4$ unit matrix, and the matrices of Clebsch-Gordan coefficients $C_{\Gamma \gamma}$ are defined by \cite{Koster1963}:
\begin{eqnarray}
 C_{E\theta} &=&
 \begin{pmatrix}
  1 & 0 & 0 & 0 \\
  0 & -1& 0 & 0 \\
  0 & 0 & -1& 0 \\
  0 & 0 & 0 & 1
 \end{pmatrix},
\quad
 C_{E\epsilon} =
 \begin{pmatrix}
  0 & 0 & 1 & 0 \\
  0 & 0 & 0 & 1 \\
  1 & 0 & 0 & 0 \\
  0 & 1 & 0 & 0
 \end{pmatrix},
\nonumber\\
 C_{T_2\xi} &=&
 \begin{pmatrix}
  0 & -i& 0 & 0 \\
  i & 0 & 0 & 0 \\
  0 & 0 & 0 & i \\
  0 & 0 & -i& 0
 \end{pmatrix},
\quad
 C_{T_2\eta} =
 \begin{pmatrix}
  0 & -1& 0 & 0 \\
  -1& 0 & 0 & 0 \\
  0 & 0 & 0 & 1 \\
  0 & 0 & 1 & 0
 \end{pmatrix},
\nonumber\\
 C_{T_2\zeta} &=&
 \begin{pmatrix}
   0 & 0 & i & 0 \\
   0 & 0 & 0 & i \\
   -i& 0 & 0 & 0 \\
   0 & -i& 0 & 0
 \end{pmatrix}.
\end{eqnarray}
The components of the $E$ representation, $\theta$ and $\epsilon$, and the $T_2$ representation, $\xi$, $\eta$ and $\zeta$,
transform as $(2z^2-x^2-y^2)/\sqrt{6}$, $(x^2-y^2)/\sqrt{2}$, $\sqrt{2} yz$, $\sqrt{2} zx$, and $\sqrt{2} xy$, respectively, under the symmetry operations.
The electronic basis of the JT Hamiltonian (\ref{Eq:HJT}) is in the increasing order of $M$.

\section{Role of the Jahn-Teller dynamics in Zeeman interaction}
\label{Sec:DJT}
We assume that the energy scale of the JT effect is larger than the Zeeman interaction.
Thus, the eigenstates of the JT Hamiltonian (vibronic states) are obtained first, and then the Zeeman interaction is rewritten in the basis of the ground vibronic state.

\begin{figure}[bt]
\begin{center}
\includegraphics[height=5cm]{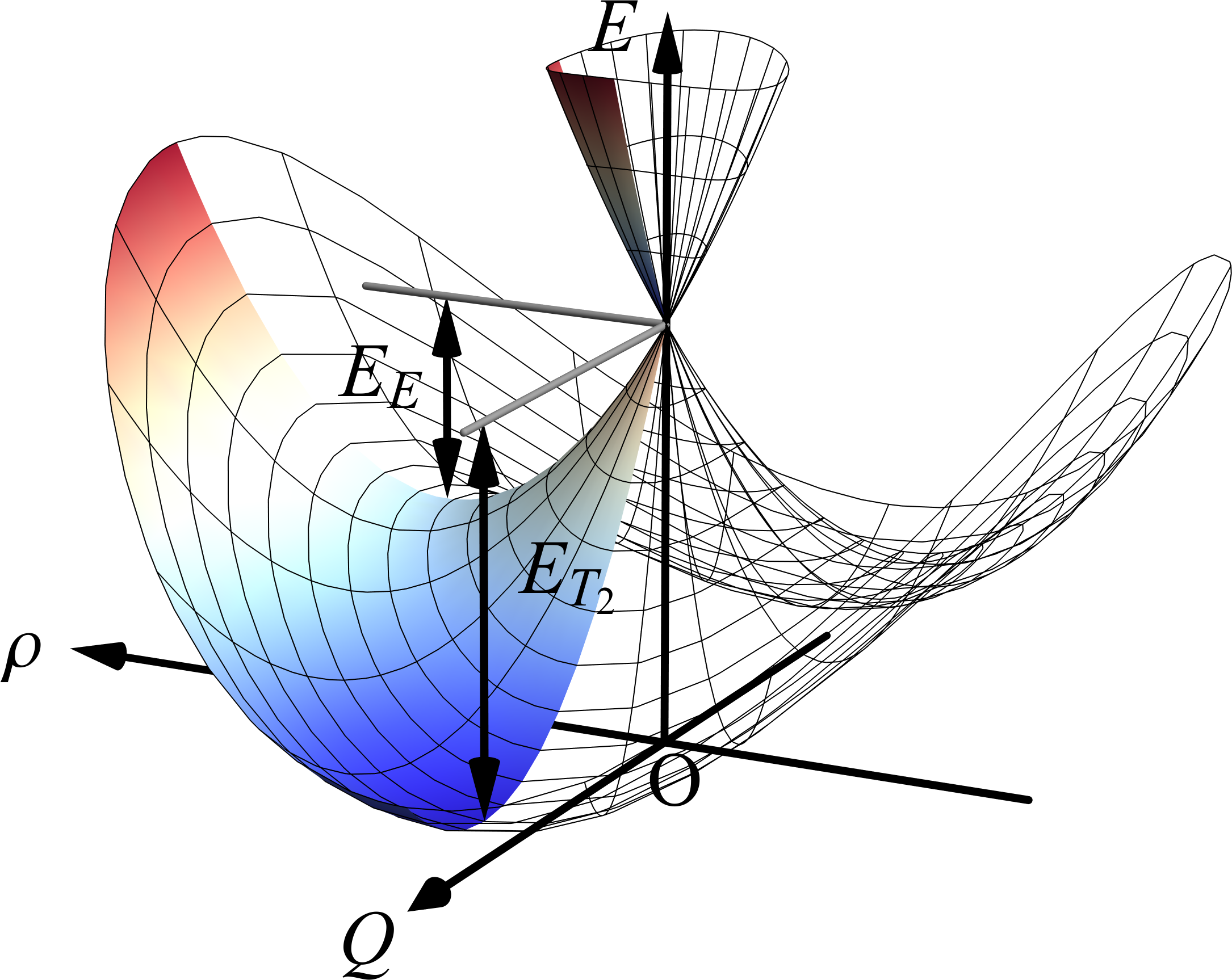}
\end{center}
\caption{(color online) The adiabatic potential energy surface of the linear $\Gamma_8 \otimes (e \oplus t_2)$ Jahn-Teller Hamiltonian, Eq. (\ref{Eq:Uad}).
The region $\rho, Q \ge 0$ is colored.
}
\label{FigC}
\end{figure}

\subsection{Adiabatic potential energy surface}
Introducing the polar coordinates,
\begin{eqnarray}
 (Q_{E\theta}, Q_{E\epsilon}) &=& \rho(\cos\phi, \sin\phi),
\nonumber\\
 (Q_{T_2\xi}, Q_{T_2\eta}, Q_{T_2\zeta}) &=& Q(\sin \alpha \cos\beta, \sin \alpha \sin \beta, \cos \alpha),
\quad
\label{Eq:polar}
\end{eqnarray}
and diagonalizing the linear vibronic term in Eq. (\ref{Eq:HJT}), we obtain the adiabatic potential energy surfaces (APES) (Fig. \ref{FigC}),
\begin{eqnarray}
 U_{\pm}(\rho, Q) &=& \frac{\omega_E^2}{2}\rho^2 + \frac{\omega_{T_2}^2}{2}Q^2 \pm \sqrt{V_E^2\rho^2+V_{T_2}^2Q^2}.
\label{Eq:Uad}
\end{eqnarray}
The domains for the polar coordinates are $\rho, Q \ge 0$, $0 \le \phi < 2\pi$, $0 \le \alpha \le \pi$, and $0 \le \beta < 2\pi$.
Under the JT distortion, the $\Gamma_8$ multiplet splits into two Kramers doublets.
The minima of the APES are expressed as
\begin{eqnarray}
 (\rho, Q, U_-) &=& \left(\frac{V_E}{\omega_E^2}, 0, -E_E\right), \left(0, \frac{V_{T_2}}{\omega_{T_2}^2}, -E_{T_2}\right),
\label{Eq:JTminima}
\end{eqnarray}
where $E_\Gamma$ is the JT stabilization energy for the $\Gamma $ vibrational mode (Fig. \ref{FigC}),
\begin{eqnarray}
 E_\Gamma = \frac{V_\Gamma^2}{2\omega_\Gamma^2} = \frac{\hslash \omega_\Gamma k_\Gamma^2}{2},
\label{Eq:EGamma}
\end{eqnarray}
and, $k_\Gamma = V_\Gamma/\sqrt{\hslash \omega_\Gamma^3}$ is the dimensionless vibronic coupling constant.

The APES (\ref{Eq:Uad}), as well as the minima (\ref{Eq:JTminima}), do not depend on the angles $\phi$, $\alpha$ and $\beta$,
which indicates the existence of the continuum of minima along these coordinates.
Thus for $E_E > E_{T_2}$ the global minima become one-dimensional continuum (trough) along $\phi$,
whereas for $E_E < E_{T_2}$ the trough is two-dimensional along $\alpha$ and $\beta$.
Finally, when $E_E = E_{T_2}$, the energy barriers between the two minima in Eq. (\ref{Eq:JTminima}) disappear, leading to a four dimensional trough.

\begin{figure*}[bt]
\begin{center}
\begin{tabular}{ccccc}
\multicolumn{1}{l}{(a)} & ~ &
\multicolumn{1}{l}{(b)} & ~ &
\multicolumn{1}{l}{(c)} \\
\includegraphics[width=5.5cm]{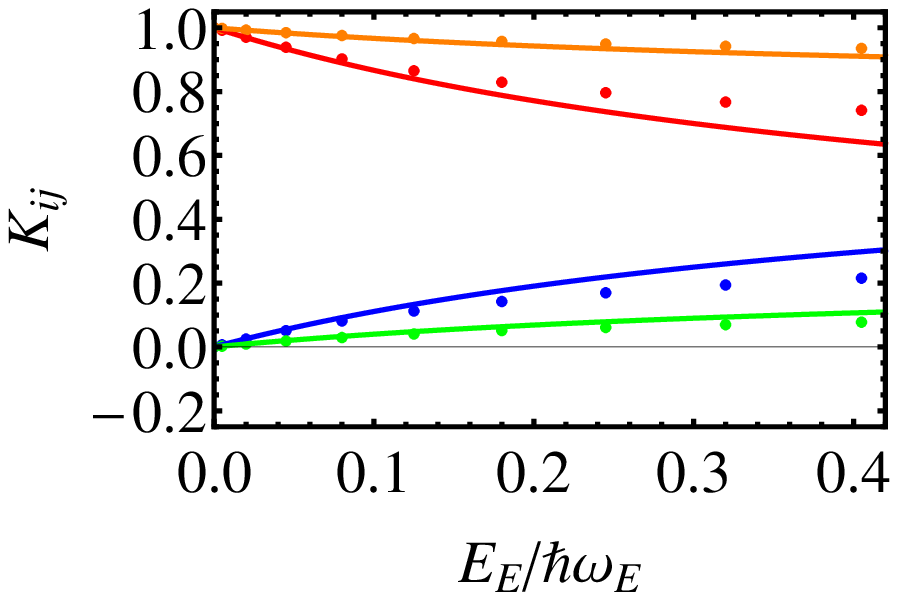}
& ~ &
\includegraphics[width=5.5cm]{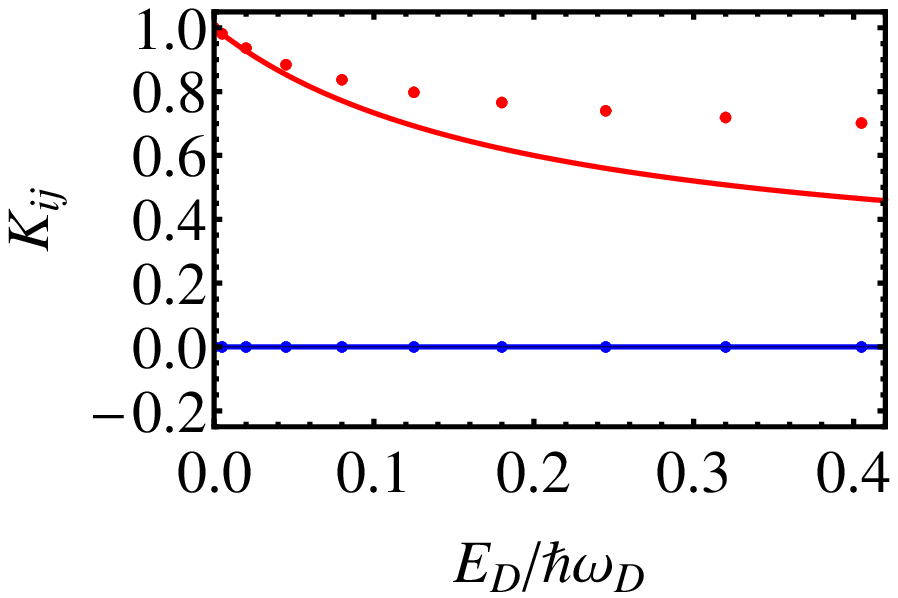}
& ~ &
\includegraphics[width=5.5cm]{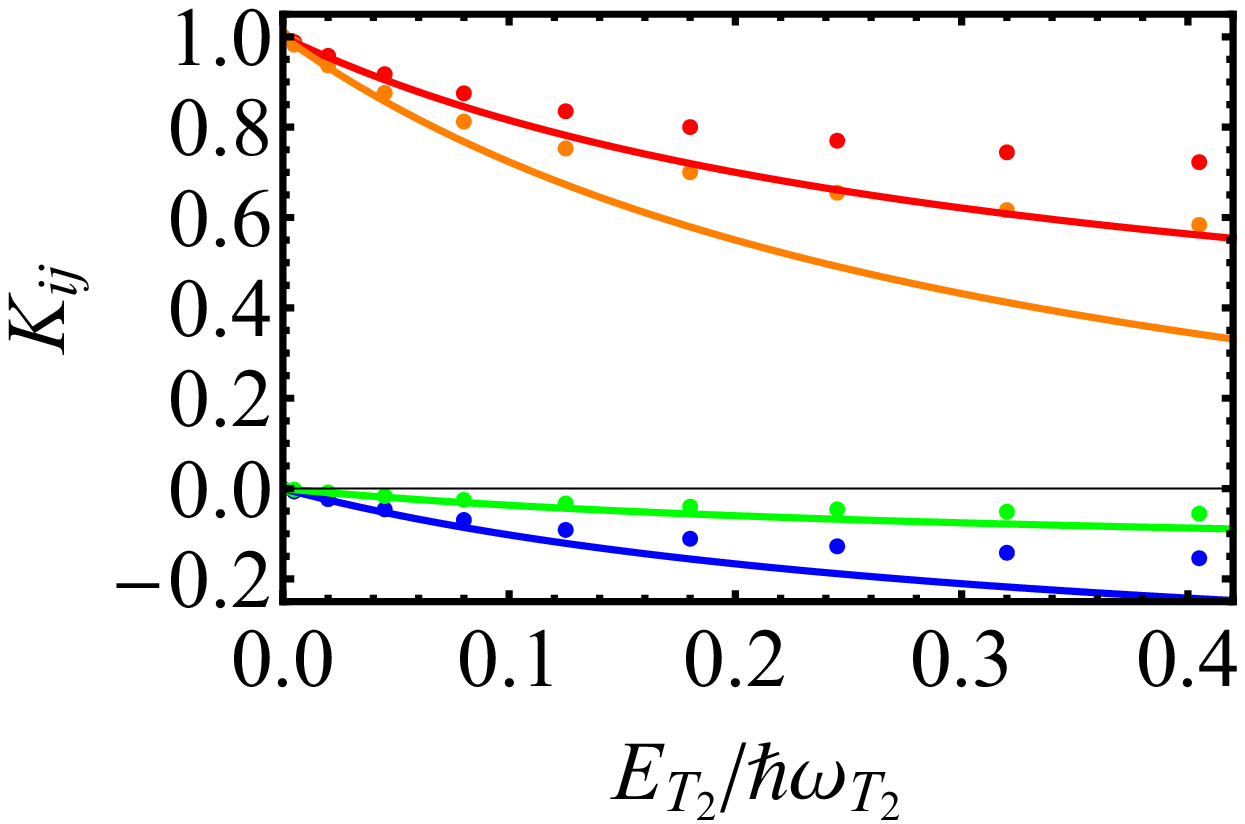}
\\
\multicolumn{1}{l}{(d)} & ~ &
\multicolumn{1}{l}{(e)} & ~ &
\multicolumn{1}{l}{(f)} \\
\includegraphics[width=5.5cm]{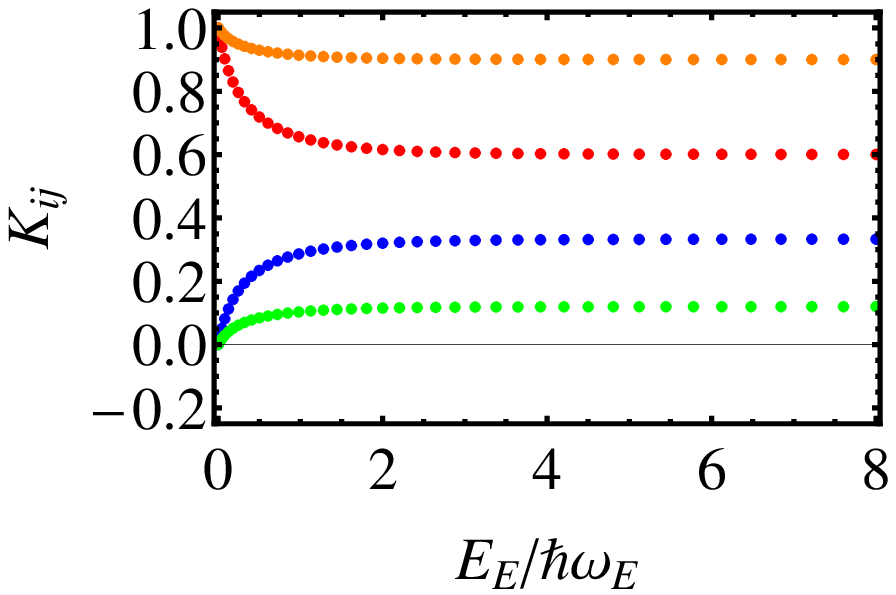}
& ~ &
\includegraphics[width=5.5cm]{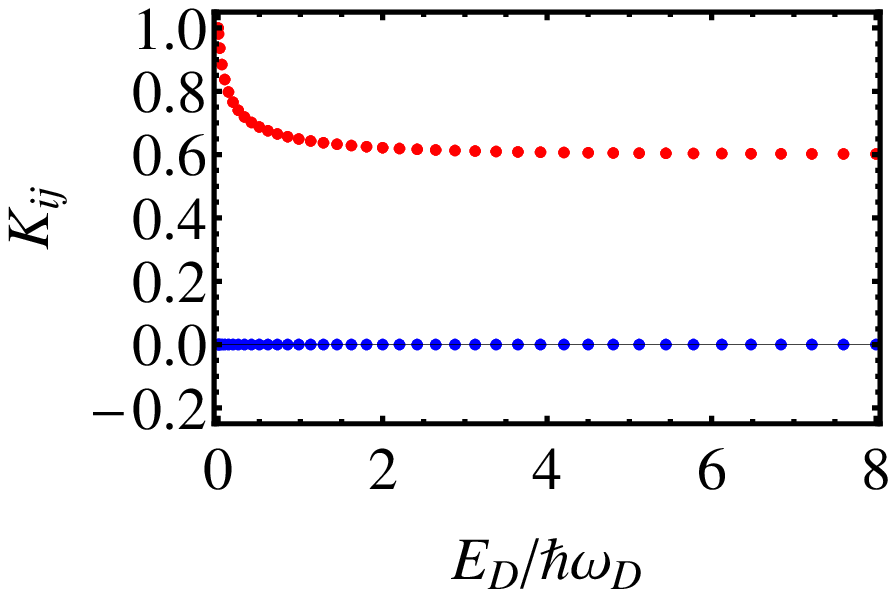}
& ~ &
\includegraphics[width=5.5cm]{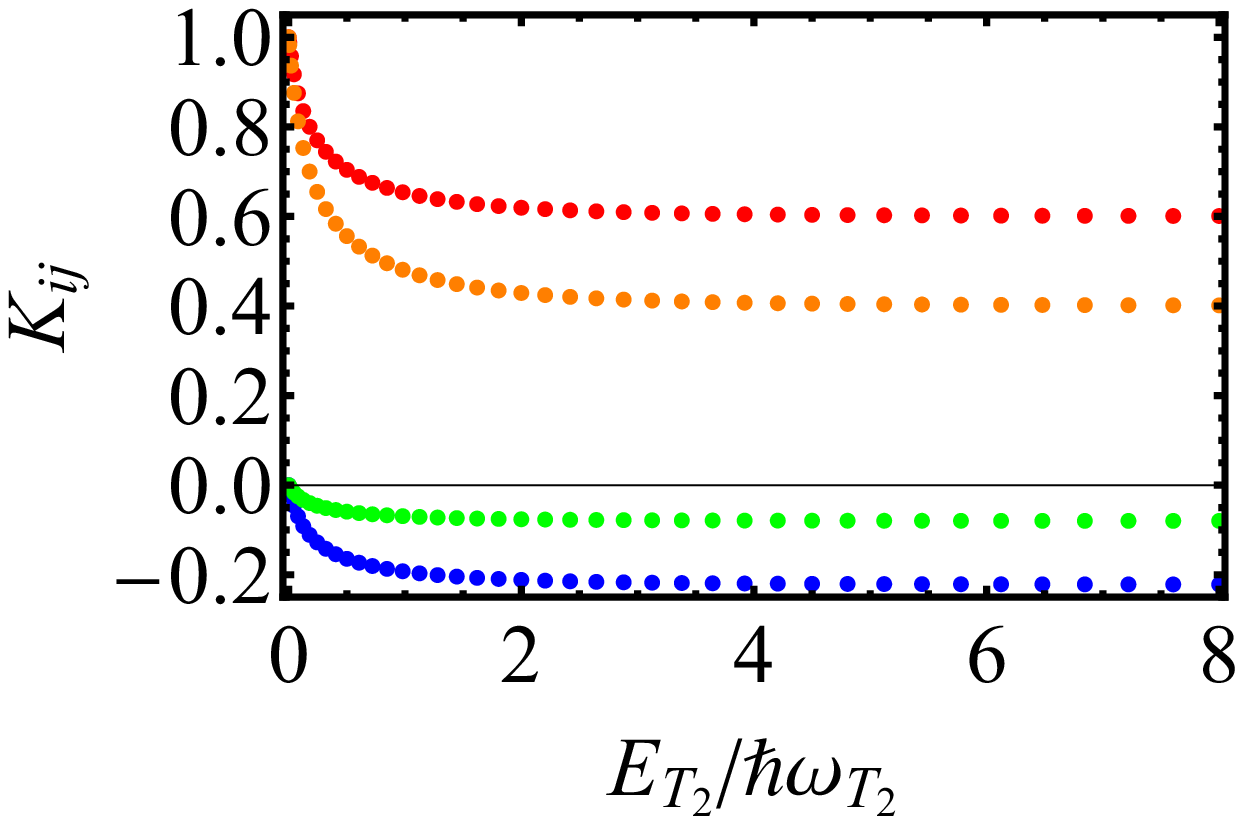}
\end{tabular}
\end{center}
\caption{(color online)
Vibronic reduction factors $K_{ij}$ as functions of the dimensionless JT stabilization energy $E_\Gamma/(\hslash \omega_\Gamma)$.
The red, orange, blue, and green symbols correspond to $K_{11}$, $K_{22}$, $K_{12}$, and $K_{21}$, respectively.
(a),(d) The $\Gamma_8 \otimes e$, (b),(e) the $\Gamma_8 \otimes d$ ($k_E = k_{T_2} = k_D, \omega_E = \omega_{T_2} = \omega_D$), and (c),(f) the $\Gamma_8 \otimes t_2$ Jahn-Teller systems.
The solid lines in (a),(b),(c) correspond to the solutions in the weak coupling limit (Eq. (\ref{Eq:K_weak})), and the bullets to the numerical solutions.
}
\label{FigE}
\end{figure*}

\begin{figure*}[tb]
\begin{center}
\begin{tabular}{ccccc}
\multicolumn{1}{l}{(a)} & ~ &
\multicolumn{1}{l}{(b)} & ~ &
\multicolumn{1}{l}{(c)} \\
\includegraphics[width=5.5cm]{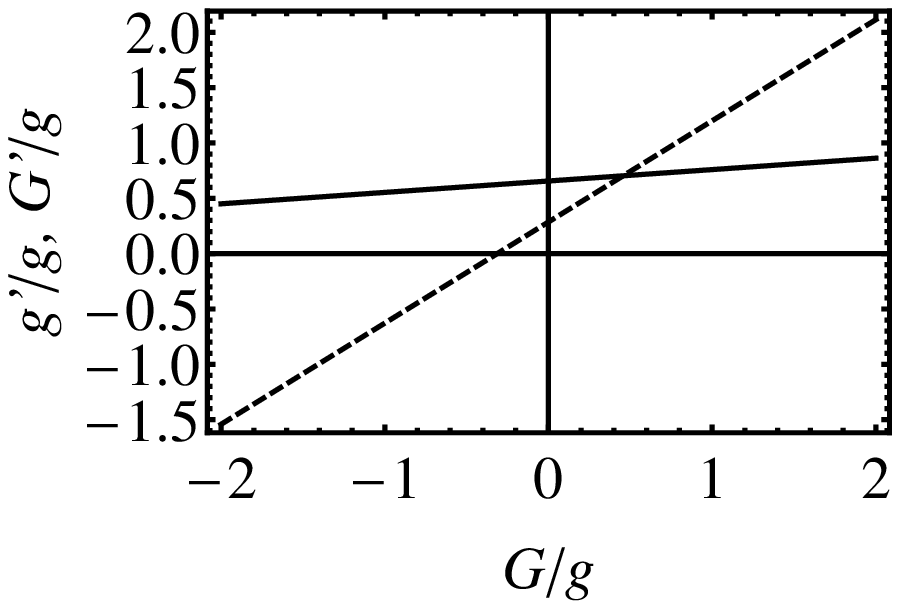}
& ~ &
\includegraphics[width=5.5cm]{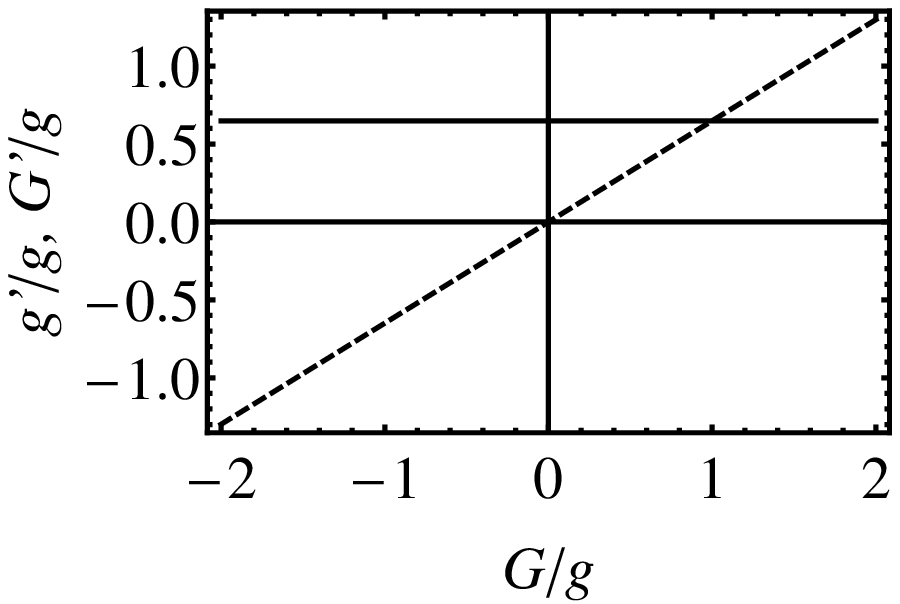}
& ~ &
\includegraphics[width=5.5cm]{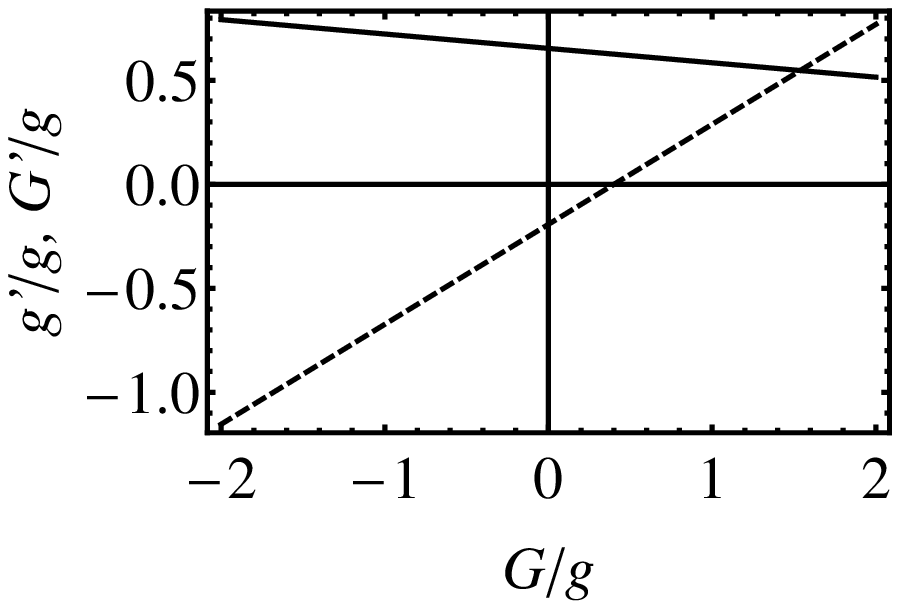}
\\
\multicolumn{1}{l}{(d)} & ~ &
\multicolumn{1}{l}{(e)} & ~ &
\multicolumn{1}{l}{(f)} \\
\includegraphics[width=5.5cm]{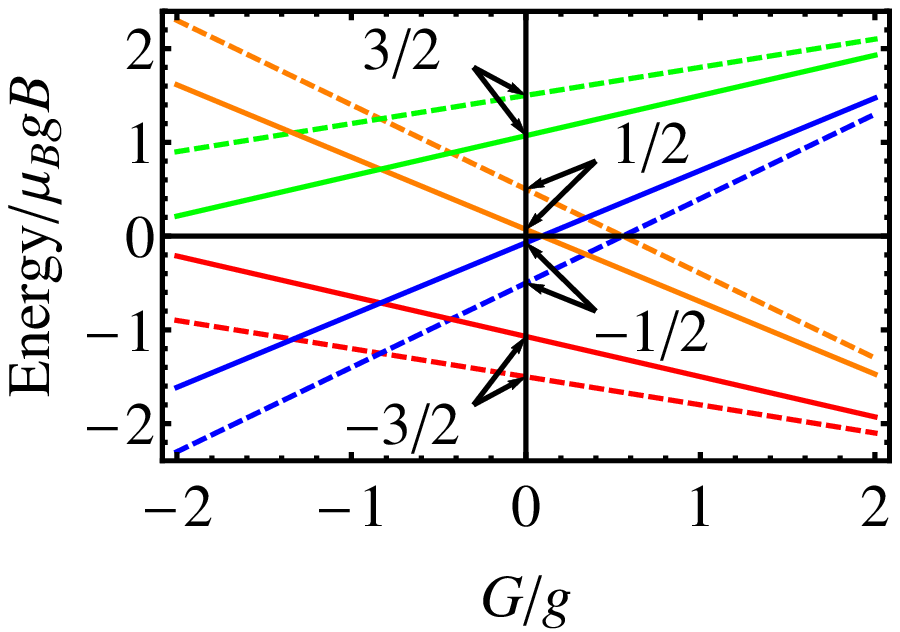}
& ~ &
\includegraphics[width=5.5cm]{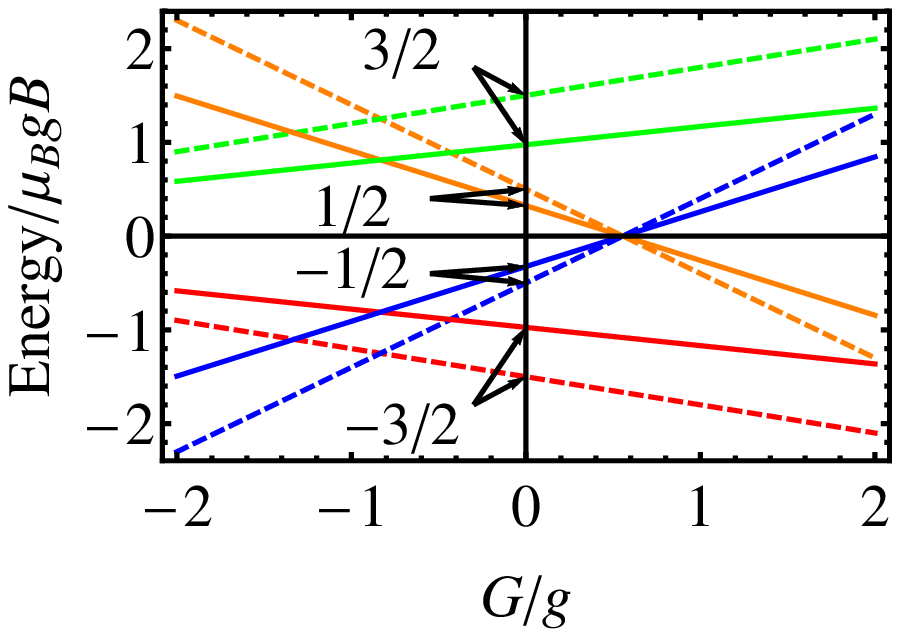}
& ~ &
\includegraphics[width=5.5cm]{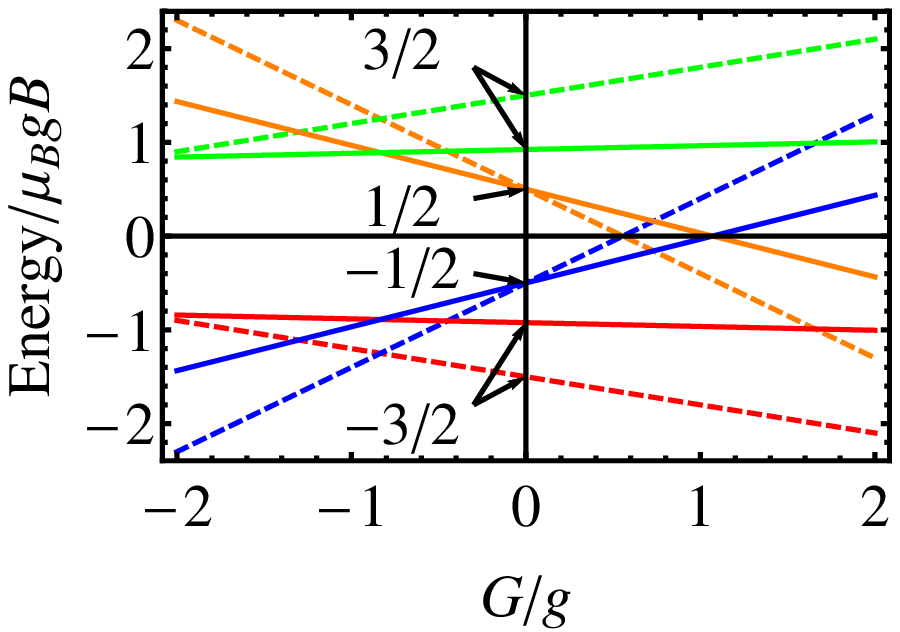}
\end{tabular}
\end{center}
\caption{(color online)
$g'$ and $G'$ for $\Gamma_8 \otimes e$ (a), $\Gamma_8 \otimes d$ (b), and $\Gamma_8 \otimes t_2$ (c) JT system.  
The ratios $g'/g$ (solid lines) and $G'/g$ (dashed lines) are given in function of $G/g$.
The energy levels of $\Gamma_8 \otimes e$ (d), $\Gamma_8 \otimes d$ (e), and $\Gamma_8 \otimes t_2$ (f) JT systems
in function of $G/g$ without (dashed lines) and with (solid lines) dynamical JT effect.
The numbers $-3/2, -1/2, 1/2, 3/2$ are momentum projection of the vibronic states.
For the calculations, the data with $k_\Gamma = 1.4$ were used, when the reduction factors are saturated.
}
\label{FigF}
\end{figure*}

\subsection{Vibronic state}
\label{Sec:Vibronic}
Because of the existence of the trough in the APES, the vibronic wave function is delocalized 
along corresponding angular mode or coordinates. 
The vibronic state is 
a superposition of the products of the electronic and vibrational wave functions:
\begin{eqnarray}
 |\Psi_{\Lambda \lambda} \rangle &=& \sum_{M=-\tilde{S}}^{\tilde{S}} |\Phi_{\Gamma_8 M}\rangle |\chi^M_{\Lambda \lambda}\rangle.
\label{Eq:vibronic_wf}
\end{eqnarray}
The vibrational part $|\chi^M_{\Lambda \lambda}\rangle$ is expanded into eigenstates of the harmonic oscillators:
\begin{eqnarray}
 |\chi^M_{\Lambda \lambda}\rangle &=&
 \sum_{n_\theta n_\epsilon n_\xi n_\eta n_\zeta}
 \chi^{M}_{\Lambda \lambda}(\bm{n})
 |n_\theta n_\epsilon n_\xi n_\eta n_\zeta\rangle,
\label{Eq:vibrational_part}
\end{eqnarray}
where, $n_\gamma$ are the number of the vibrational quanta (indices $E$ and $T_2$ are omitted for simplicity); 
$\bm{n} = (n_\theta, n_\epsilon, n_\xi, n_\eta, n_\zeta)$, and $\chi^{M}_{\Lambda \lambda}(\bm{n})=\langle \bm{n}|\chi^M_{\Lambda \lambda}\rangle $ are decomposition coefficients. 
According to the general rule, the irreducible representations of the electronic state and
the ground vibronic state $\Lambda$ in linear JT systems coincide \cite{Bersuker1989}.
Thus, in the present case $\Lambda = \Gamma_8$.
Using the $\Gamma_8$ ground vibronic states, $\tilde{S}=3/2$ pseudospin 
can be defined following the rules discussed in the previous section.

As an example of the vibronic state (\ref{Eq:vibronic_wf}), the expression in the weak vibronic coupling limit ($\hslash \omega_\Gamma \gg E_\Gamma$) within first order perturbation is given in Appendix \ref{Sec:Vibronic_pert}.
We stress that the vibronic state is not a simple product of the electronic and the vibrational states as in the case of the nondegenerate system (Born-Oppenheimer approximation), but represents an entangled electron-nuclear state.
In the strong coupling case ($E_\Gamma \gg \hslash \omega_\Gamma$), the ground vibronic wave function is well described by
the product of the ground adiabatic electronic state, the radial vibrational wave function ($\rho, Q$), and the rotational wave function in the trough ($\alpha, \beta, \phi$) \cite{Bersuker1989, Bersuker2006}.
For the details of the linear $\Gamma_8 \otimes (e \oplus t_2)$ JT system in the strong vibronic coupling limit, see e.g. Ref. \cite{Apsel1992}.

In most of the existing systems, the strength of the vibronic coupling is intermediate ($E_\Gamma \approx \hslash \omega_\Gamma$).
Accurate vibronic states can be only obtained by numerical diagonalization of the JT Hamiltonian matrix (\ref{Eq:HJT}).
In our numerical calculations, the vibrational basis $\{|n_\theta n_\epsilon n_\xi n_\eta n_\zeta\rangle\}$ in Eq. (\ref{Eq:vibrational_part}) was truncated,
$0 \le \sum_\gamma n_\gamma \le 40$, and Lanczos algorithm was used for the diagonalization.

\subsection{Vibronic reduction factor}
\label{Sec:VibronicRed}
Within the space of the ground vibronic states, the electronic operators are modified by vibronic reduction factors (Ham factors) \cite{Englman1972, Bersuker1989, Ham1976}.
Particularly, as seen in Eq. (\ref{Eq:selection1}), the double cubic group is not simply reducible, and each of $t_1$ operators is expressed by two parameters.
Following Ref. \cite{Ham1976}, we introduce the vibronic reduction factors $K_{ij}$ for $t_1 (\Gamma_4)$ operators as
\begin{eqnarray}
 \langle \Psi_{\Gamma_8M}|O_1^z(\tilde{\bm{S}})|\Psi_{\Gamma_8M}\rangle
 &=&
  K_{11} O_1^z(M) + K_{12} O_3^z(M),
\quad
\nonumber \\
 \langle \Psi_{\Gamma_8M}|O_3^z(\tilde{\bm{S}})|\Psi_{\Gamma_8M}\rangle
 &=&
  K_{21} O_1^z(M) + K_{22} O_3^z(M),
\quad
\label{Eq:Kij}
\end{eqnarray}
where, $O_k^z(M) = \langle \Phi_{\Gamma_8M}|O_k^z(\tilde{\bm{S}})|\Phi_{\Gamma_8M}\rangle$.

In the weak vibronic coupling limit (Appendix \ref{Sec:Vibronic_pert}), the reduction factors are calculated as
\begin{eqnarray}
 K_{11} &=& \frac{5+k_E^2+3k_{T_2}^2/2}{5(1+k_E^2+3k_{T_2}^2/2)},
\nonumber\\
 K_{12} &=& \frac{2(k_E^2-k_{T_2}^2)}{3(1+k_E^2+3k_{T_2}^2/2)},
\nonumber\\
 K_{21} &=& \frac{9}{25}K_{12},
\nonumber\\
 K_{22} &=& \frac{5+4k_E^2-3k_{T_2}^2/2}{5(1+k_E^2+3k_{T_2}^2/2)}.
\label{Eq:K_weak}
\end{eqnarray}
The reduction factors significantly depend on the type of the Jahn-Teller effect.
When $|k_E| > |k_{T_2}|$, we obtain $K_{11} < K_{22}$ and $K_{12}, K_{21} > 0$, while when $|k_E| < |k_{T_2}|$, $K_{11} > K_{22}$ and $K_{12}, K_{21} < 0$.
These relations for $K_{12}$ and $K_{21}$ hold for any strength of the vibronic coupling \cite{Ham1976}.
The solid lines in Figure \ref{FigE} (a)-(c) are calculated reduction factors in Eq. (\ref{Eq:K_weak}) with respect to the vibronic coupling strength for various situations.

Fig. \ref{FigE} (e) shows good agreement with the simulation in Ref. \cite{OBrien1979}.
In comparison to numerical results, the reduction factors within the perturbation theory are quantitatively correct only for
$E_{\Gamma}/\hslash \omega_\Gamma \alt 0.1$, while their qualitative properties such as the sign and the order of $K_{ij}$ are well described
even for larger $E_\Gamma$ (Fig. \ref{FigE} (a)-(c)).
Figure \ref{FigE}(d)-(f) shows that the reduction factors are saturated around $E_\Gamma/\hslash \omega_\Gamma \approx 1$.

\subsection{$g$ factors}
\label{Sec:gfactor}
In the ground vibronic states, the magnetic moment operators are calculated as follows:
\begin{eqnarray}
 \mu'_\alpha &=&
 -\mu_{\rm B} g' O_1^\alpha(\tilde{\bm{S}})
 -\mu_{\rm B} G' O_3^\alpha(\tilde{\bm{S}}),
\label{Eq:mu_vibronic}
\end{eqnarray}
where, $g'$ and $G'$ are defined by
\begin{eqnarray}
 g' = K_{11}g +K_{21}G,
\quad
 G' = K_{12}g + K_{22}G,
\label{Eq:gG_DJT}
\end{eqnarray}
respectively, using Eq. (\ref{Eq:Kij}). 
The pseudospin operator $\tilde{\bm{S}}$ in Eq. (\ref{Eq:mu_vibronic}) acts on the ground $\Gamma_8$ vibronic states. 
As Eq. (\ref{Eq:mu_vibronic}) shows, the Zeeman pseudospin Hamiltonian remains isotropic as in Eq. (\ref{Eq:mu_Gamma8}).

Figure \ref{FigF}(a)-(c) shows $g'/g$ and $G'/g$ in function of $G/g$ for $\Gamma_8 \otimes e$, $\Gamma_8 \otimes d$, and $\Gamma_8 \otimes t_2$ models, respectively.
Because of the reduction factors $K_{ij}$, $g'$ and $G'$ can change their signs \cite{Halliday1988}.
In particular, for the $\Gamma_8 \otimes e$ and the $\Gamma_8 \otimes t_2$ JT systems, there are regions of $G/g$ where the sign of $G'/g$ becomes opposite to $G/g$.
Figure \ref{FigF} (d)-(f) shows the Zeeman splitting in function of $g/G$.
The difference between the Zeeman splittings by electronic $g, G$ (dashed lines) and those by vibronic $g', G'$ (solid lines) 
is evidently significant.
The Zeeman splittings are significantly modified by the JT dynamics, and the crossing point ($G/g$) of the Zeeman levels are also shifted due to the JT effect.
Thus, the interpretation of EPR and related experiments based solely on the electronic multiplets can be incorrect.

\begin{figure*}[tb]
\begin{center}
\begin{tabular}{ccccc}
\multicolumn{1}{l}{(a)} & ~ &
\multicolumn{1}{l}{(b)} & ~ &
\multicolumn{1}{l}{(c)} \\
\includegraphics[width=5.5cm]{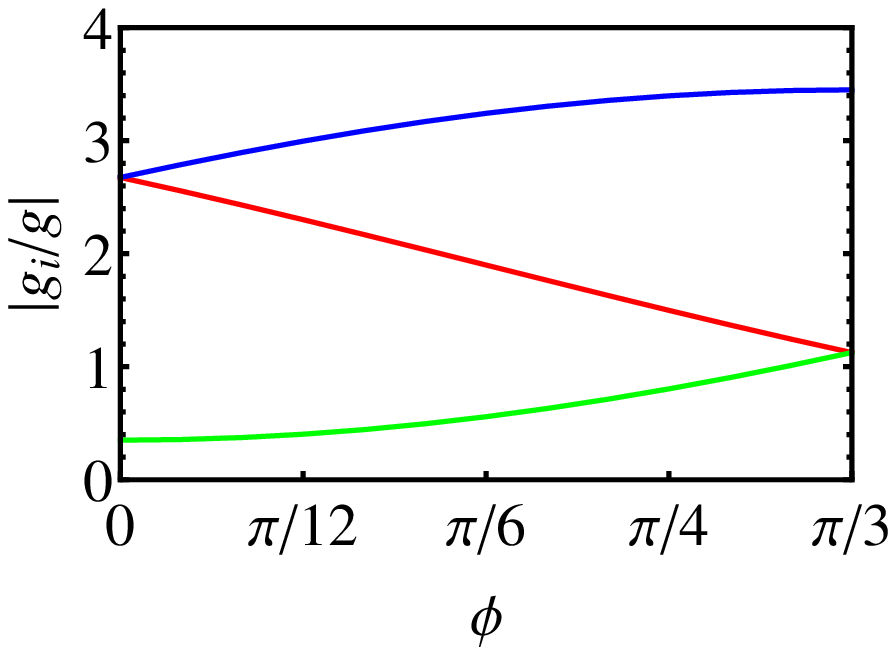}
& ~ &
\includegraphics[width=5.5cm]{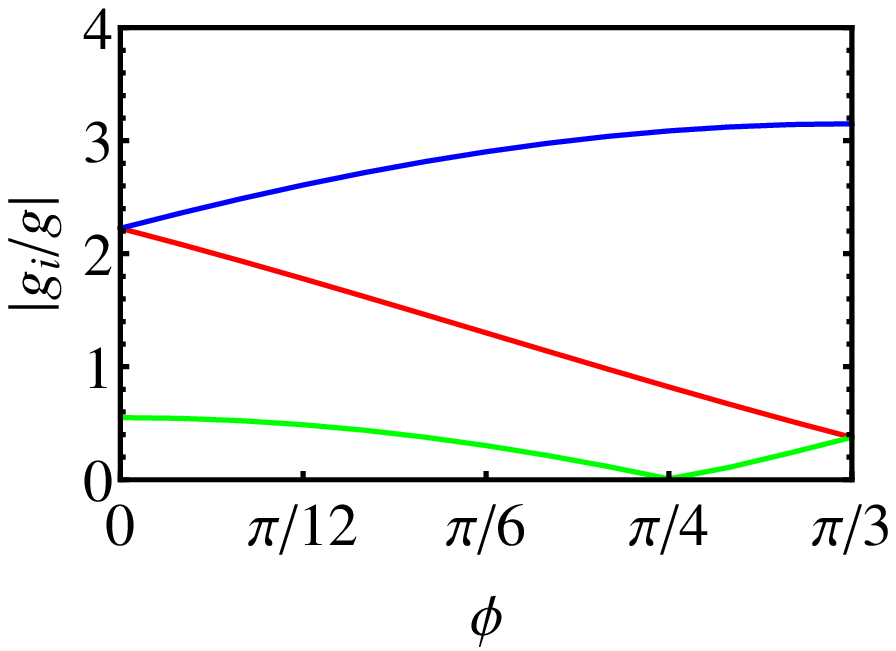}
& ~ &
\includegraphics[width=5.5cm]{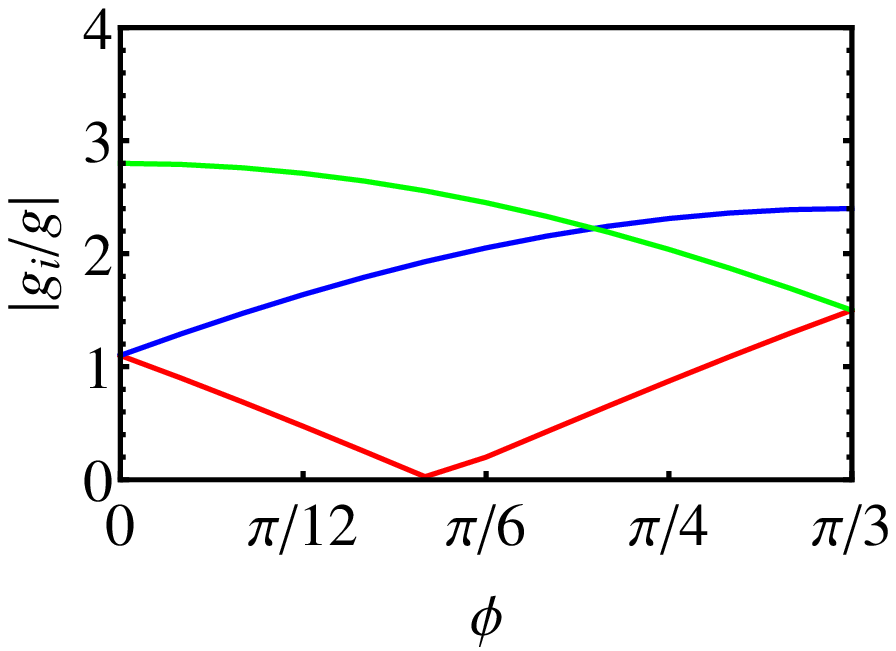}
\\
\multicolumn{1}{l}{(d)} & ~ &
\multicolumn{1}{l}{(e)} & ~ &
\multicolumn{1}{l}{(f)} \\
\includegraphics[width=5.5cm]{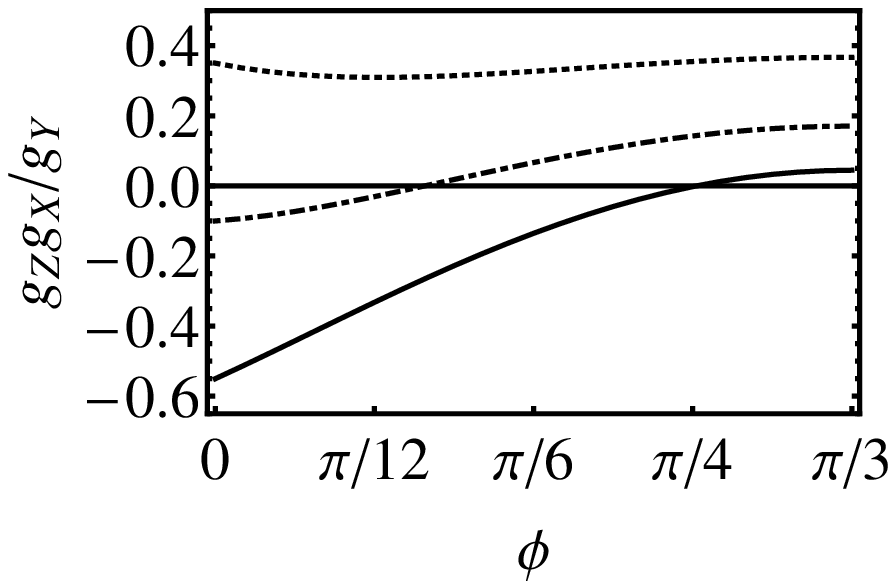}
& ~ &
\includegraphics[width=5.5cm]{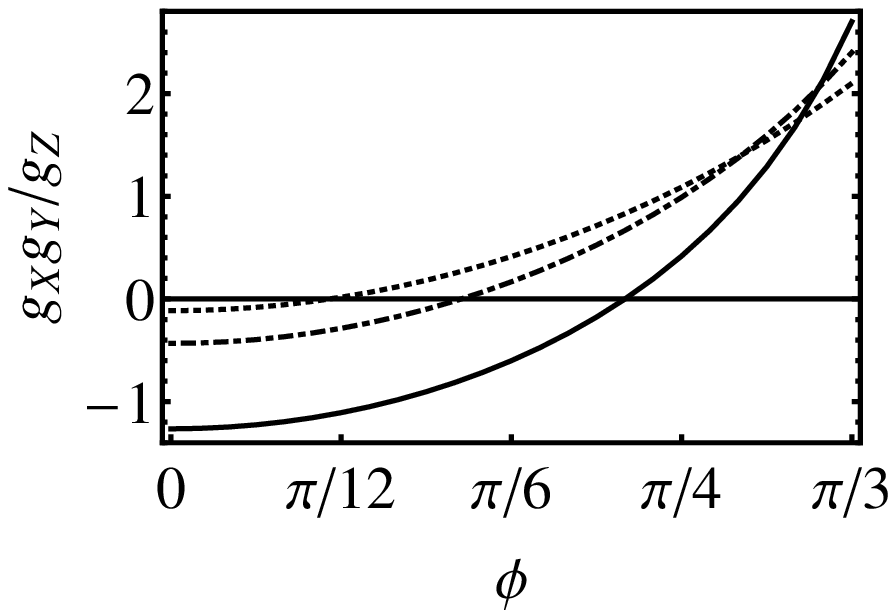}
& ~ &
\includegraphics[width=5.5cm]{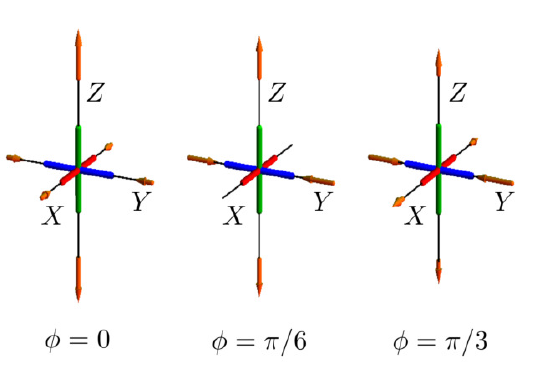}
\end{tabular}
\end{center}
\caption{(color online)
(a-c) $g_i$ $(i = X, Y, Z)$, (d,e) the signs of $g_Xg_Yg_Z$, and (f) the main magnetic axes
as functions of the angle $\phi$ of the static $e$ JT distortion.
$g_i$'s are calculated in the unit of $g$.
For the calculations of the $g$ factors, (a) $G/g = 3/4$, (b) $1/4$, and (c) $-1$ are used.
(d) $g_Zg_X/g_Y$ for $G/g = 1/4, 1/2, 3/4$
(solid, dot dashed, and dotted curves, respectively), and
(e) $g_Xg_Y/g_Z$ for $G/g = -1/2, -1, -3/2$
(solid, dot dashed, and dotted curves, respectively).
The orange arrows in plot (f) indicate the JT displacements of atoms.
}
\label{FigG}
\end{figure*}

\begin{figure*}[tb]
\begin{center}
\begin{tabular}{ccccc}
\multicolumn{1}{l}{(a)} & ~ &
\multicolumn{1}{l}{(b)} & ~ &
\multicolumn{1}{l}{(c)} \\
\includegraphics[width=5.5cm]{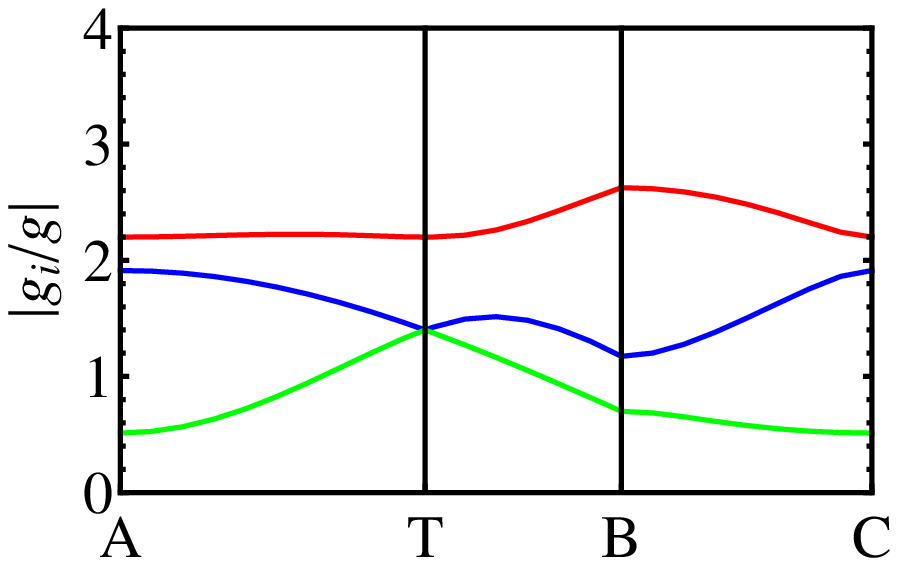}
& ~ &
\includegraphics[width=5.5cm]{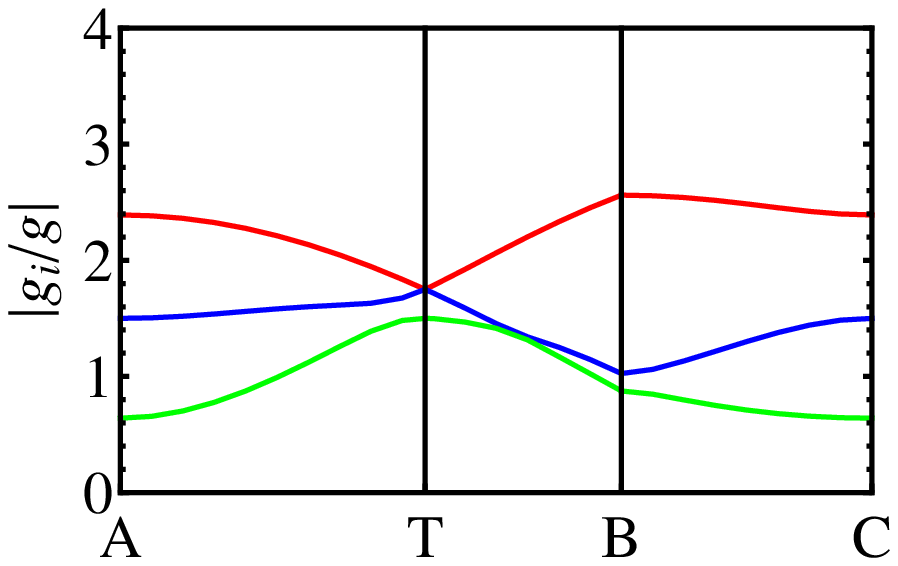}
& ~ &
\includegraphics[width=5.5cm]{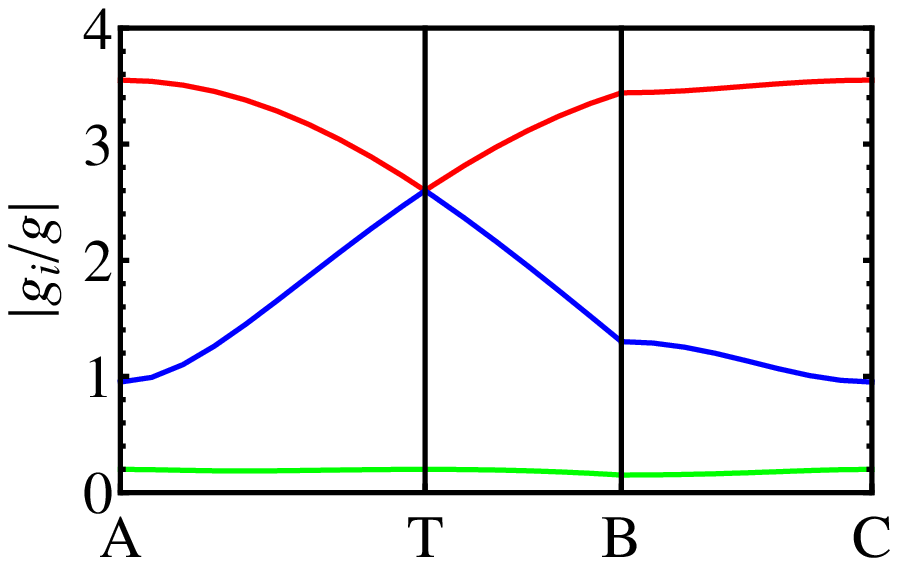}
\\
\multicolumn{1}{l}{(d)} & ~ &
\multicolumn{1}{l}{(e)} & ~ &
\multicolumn{1}{l}{(f)} \\
\includegraphics[width=5.5cm]{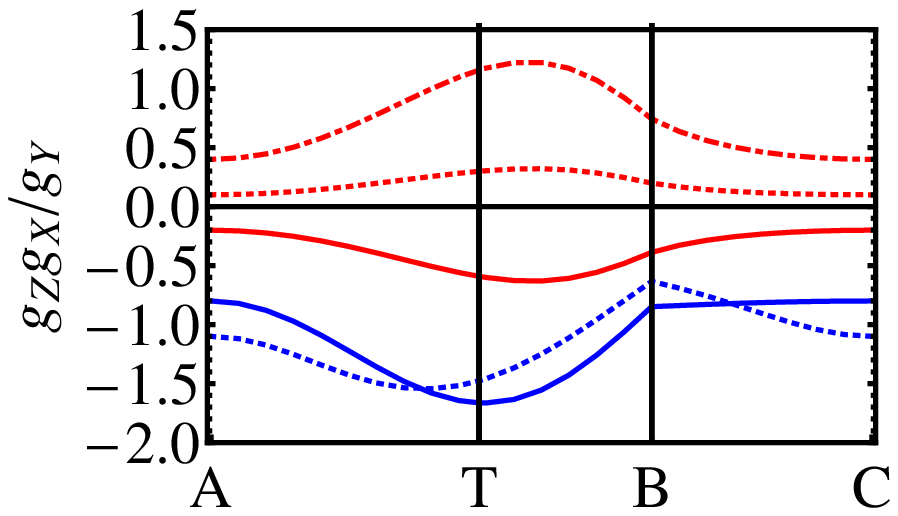}
& ~ &
\includegraphics[width=5.5cm]{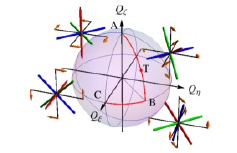}
& ~ &
\includegraphics[width=5.5cm]{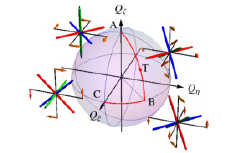}
\end{tabular}
\end{center}
\caption{(color online)
(a-c) $g_i$ $(i = X, Y, Z)$, (d) the product of the $g$ factors, and (e,f) the magnetic axes,
as functions of the Euler angles of the static $t_2$ JT distortion.
$g_i$'s are calculated in the unit of $g$.
For the calculations of the $g$ factors, (a) $G/g = 1$, (b) $5/12$, and (c) $-1$ were used.
(d) $g_Zg_X/g_Y$ for $G/g=-3/2, -1, -1/2, 1/2, 1$ (red dot-dashed, red dotted, red solid, blue solid and blue dotted, respectively) are shown.
The magnetic axes are obtained for (e) $G/g = 1$ and (f) $-1$.
The $T$ (trigonal) point corresponds to $Q_\xi = Q_\eta = Q_\zeta$.
}
\label{FigH}
\end{figure*}

\section{Role of the static Jahn-Teller distortion in Zeeman interaction}
\label{Sec:SJT}
When the higher order vibronic couplings are also important and the trough is strongly warped to mix several low-energy vibronic states, 
the ground vibronic state is localized at minima, i.e., static JT effect arises.
Let us consider the case where the splitting of the $\Gamma_8$ states is large compared with the Zeeman splitting of each Kramers doublets.
In such cases, it is convenient to introduce the pseudospin $\tilde{S}=1/2$ for each Kramers doublet
following the methodology described in Refs. \cite{Chibotaru2012JCP, Chibotaru2013}.

The magnetic moment within the lower adiabatic electronic states is
$\bm{\mu}'' = \sum_{r,r'=1}^2 \langle \Phi_r|\bm{\mu}|\Phi_{r'}\rangle |\Phi_r\rangle \langle \Phi_{r'}|$,
where $|\Phi_r(\Omega)\rangle$ $(r = 1, 2)$ are the components of the ground Kramers doublet, and $\Omega = (\phi, \alpha, \beta)$ is the set of the angles in Eq. (\ref{Eq:polar}).
Since $|\Phi_r\rangle$ depends on the JT distortion $\Omega$, $\bm{\mu}''$ and $g_i$ are also functions of $\Omega$.

Figure \ref{FigG} (a)-(c) show $g_i$ ($i=X,Y,Z$) 
as function of $\phi$ at various $G/g$ for the case $E_E > E_{T_2}$.
Due to the JT distortion, the $g$ tensor becomes anisotropic, whereas the main magnetic axes in the distorted geometries 
are directed along the tetragonal axes (Fig. \ref{FigG} (f)) since the JT distortion along $e$ modes preserve them as symmetry axes. 
Even for fixed $G/g$, e.g. $G/g = 3/4$ (Fig. \ref{FigG} (a)), the JT distortion can change the nature of the magnetic moment from
easy-plane ($g_X = g_Y > g_Z$ at $\phi = 0$) to axial ($g_Y > g_X = g_Z$ at $\phi = \pi/3$).

It is interesting to see the evolution of the sign of product $g_X g_Y g_Z$,
defining the direction of Larmor precession of the magnetic moment, the pattern of the hyperfine splitting (nuclear quadrupole moment) \cite{Pryce1959},
and the sign of the Berry phase \cite{Chibotaru2012PRL}.
The sign of $g_X g_Y g_Z$ can be found by calculating $g_i g_j/g_k$ from the expression \cite{Pryce1959}:
\begin{eqnarray}
 [\mu''_i, \mu''_j] &=& -i \mu_{\rm B} \frac{g_ig_j}{g_k} \mu''_k,
\label{Eq:ggg}
\end{eqnarray}
where, $(i,j,k)$ is a cyclic permutation of $(X,Y,Z)$. 
$g_ig_j/g_k$ is shown in Fig. \ref{FigG} (d), (e) in function of the angle $\phi$.
The calculation shows that the sign changes for some $G/g$.
The change of sign occurs at a value of $\phi$ when one of the $g$ factors become zero at a point (Fig. \ref{FigG}(b) and (c)).
Thus, the sign of the product of the $g$ factors is sensitive to the local JT distortion as well as the nature of the $\Gamma_8$ electronic states via the ratio $G/g$.

Figure \ref{FigH}(a)-(c) show $g_i$ $(i=X,Y,Z)$ in functions of the $t_2$ JT distortions, angular coordinates $\alpha$ and $\beta$ for the case $E_{T_2} > E_E$.
Contrary to the $\Gamma_8 \otimes e$ JT system, the sign of the product $g_Xg_Yg_Z$ do not change with respect to the JT distortion (Fig. \ref{FigH}(d)).
In the case of the $\Gamma_8 \otimes t_2$ JT system, however, both the values of $g_i$ and the directions of the main magnetic axes vary with distortions (Fig. \ref{FigH} (e) and (f) for $G/g=1$ and $-1$, respectively).
The nature of the main magnetic axes $g_i$ depends on the value of $G/g$. 
Thus, for $G/g = 1$ the anisotropy is of easy axis type (Fig. \ref{FigH} (a), (e)), whereas it is of easy plane type for $G/g = -1$ (Fig. \ref{FigH} (c), (f)).

\section{{\it Ab initio} derivation of $\Gamma_8$ Hamiltonian}
\label{Sec:abinitio}
As examples illustrating the obtained results, $g$ and $G$ factors and vibronic couplings of Cs$_2$ZrCl$_6$:Np$^{4+}$ impurity in octahedral zirconium site \cite{Bray1978, Bernstein1979, Edelstein1980} and of ThO$_2$:Ir$^{4+}$ impurity in cubic thorium site are further calculated.

\subsection{Computational methodology}
The low-lying electronic states of impurity are obtained from cluster calculations.
The electronic structures are calculated combining the complete active space self-consistent field (CASSCF),
XMS-CASPT2 \cite{Granovsky, Shiozaki} (for Ir$^{4+}$), and spin-orbit restricted active space state interaction (SO-RASSI) methods implemented in
 {\tt Molcas} package \cite{Molcas8}.
The cluster consists of one impurity ion surrounded by several layers of ions.
The impurity and the closest atoms are treated {\it ab initio}, whereas the others are replaced by embedding {\it ab initio} model potentials (AIMP).

\begin{figure}[tb]
\includegraphics[width=8.0cm]{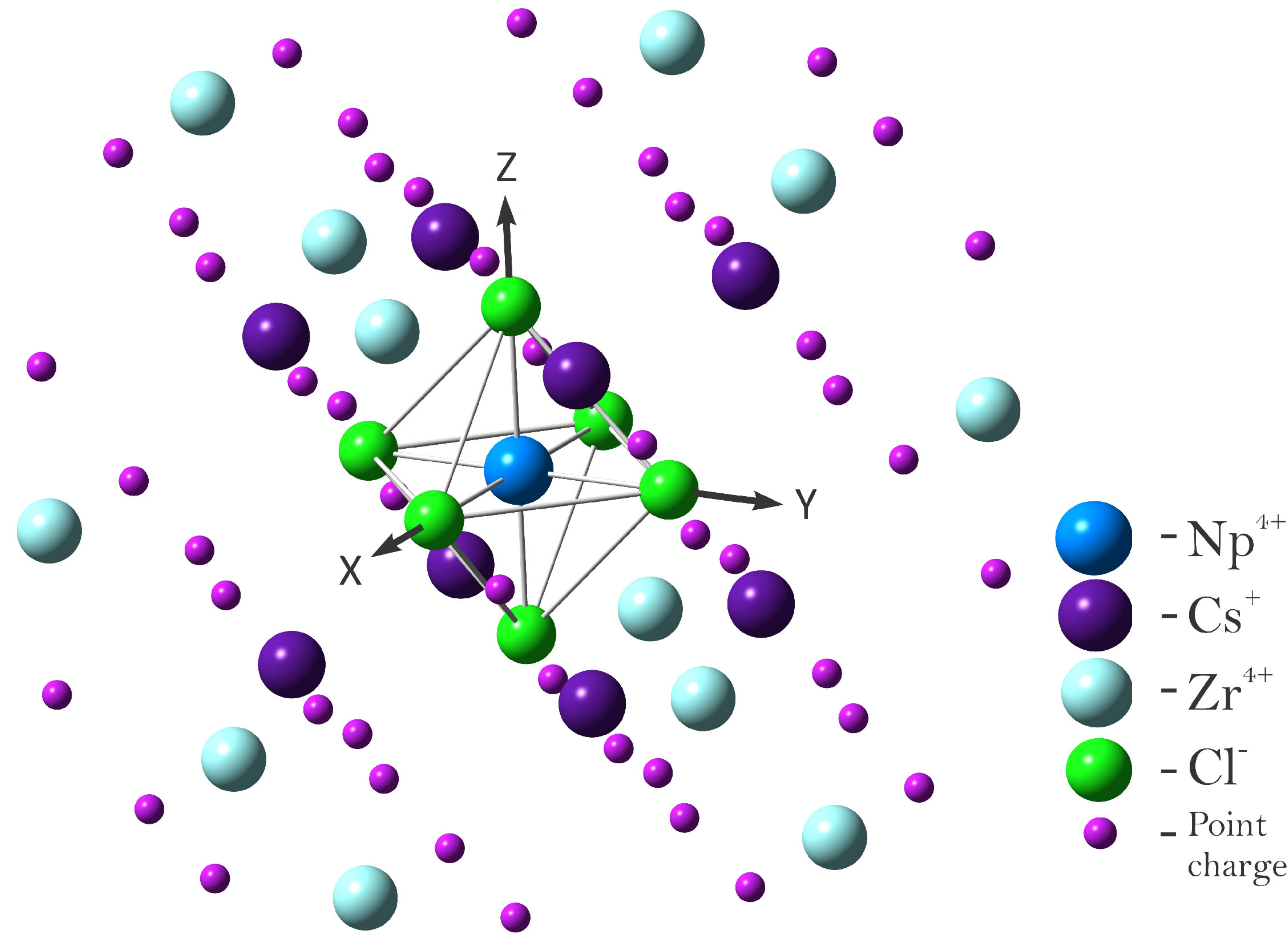}
\caption{
(color online)
The calculated Cs$_2$ZrCl$_6$:Np$^{4+}$ cluster (see the text for details).
}
\label{FigI}
\end{figure}

In the former system, Np$^{4+}$ ion and the closest six chlorine and eight cesium ions are treated {\it ab initio} (Fig. \ref{FigI}).
ANO-RCC-MB (MB), ANO-RCC-VDZP (DZP), and ANO-RCC-VTZP (TZP) basis sets were used for the Np$^{4+}$ and closest chlorine ions and effective core potential (ECP)-AIMP
was used for the closest eight cesium ions \cite{Zoila-Cs-ECP} which included the last seven electrons in the basis.
Contraction of the employed basis sets are given in the Supplemental Materials \cite{SM}.
The next two layers of atoms are described by {\it ab initio} model potentials (AIMP) with no electrons (frozen density).
Optimization of these potentials were done within the iterative self-consistent embedding ion (SCEI) methodology 
\cite{Seijo_SCEI} employed earlier.
For these calculations we used similar size basis sets for individual atoms. In this approach, {\it ab initio} self consistent calculations are carried out for
each atom in the field of model potentials of other atoms. The SCEI procedure is carried out until full self consistency between potentials of all
atoms is achieved (${\Delta}E < 10^{-7}$ a.u.). Optimized AIMPs for the individual atoms in Cs$_{\text 2}$ZrCl$_{\text 6}$ are given in \cite{SM}.
The active space consisted of three electrons in seven 4$f$-type orbitals. All spin-free states (35 quartet and 112 doublet states)
were mixed by spin-orbit coupling within RASSI program.

In the latter system, Ir$^{4+}$ ion and the closest eight oxygen atoms were treated {\it ab initio} by employing
ANO-RCC-VDZP basis set. 
For the distant thorium and oxygen atoms, the Ba-ECP-AIMP \cite{Seijo-Ba-ECP} and O-ECP-AIMP \cite{Pascual-O-ECP} were used, respectively.
The active space comprised eleven electrons in 8 orbitals (five 5$d$-type orbitals of Ir$^{4+}$ and three $t_{2g}$-type orbitals of ligand).
75 doublet states ($S=1/2$) were optimized at the CASSCF level. 
In the multi-state-XMS-CASPT2 calculation only the lowest three roots were considered, which were later mixed by spin-orbit coupling at SO-RASSI stage.

\begin{figure*}[tb]
\begin{center}
\begin{tabular}{ccccccc}
\multicolumn{1}{l}{(a)} & & 
\multicolumn{1}{l}{(b)} & &
\multicolumn{1}{l}{(c)} & &
\multicolumn{1}{l}{(d)} \\
\includegraphics[height=3.3cm]{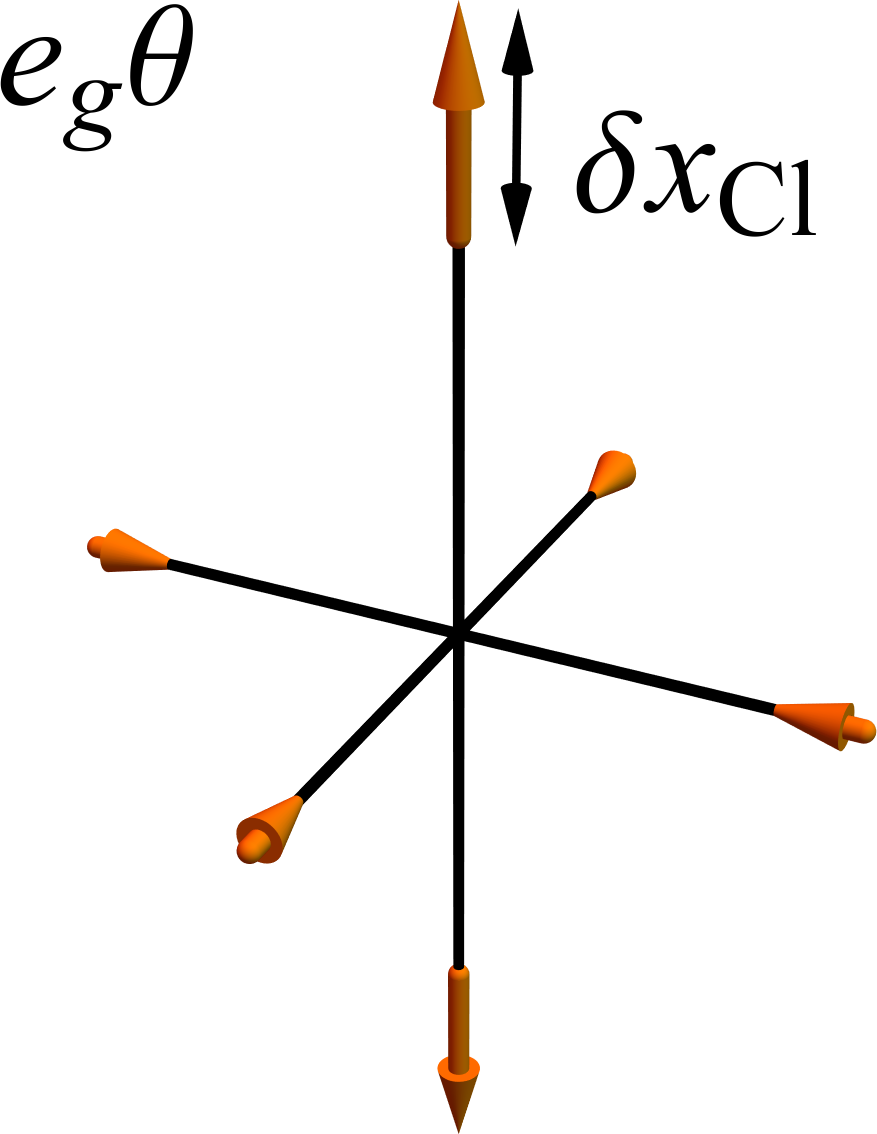}
&
~
&
\includegraphics[height=3.3cm]{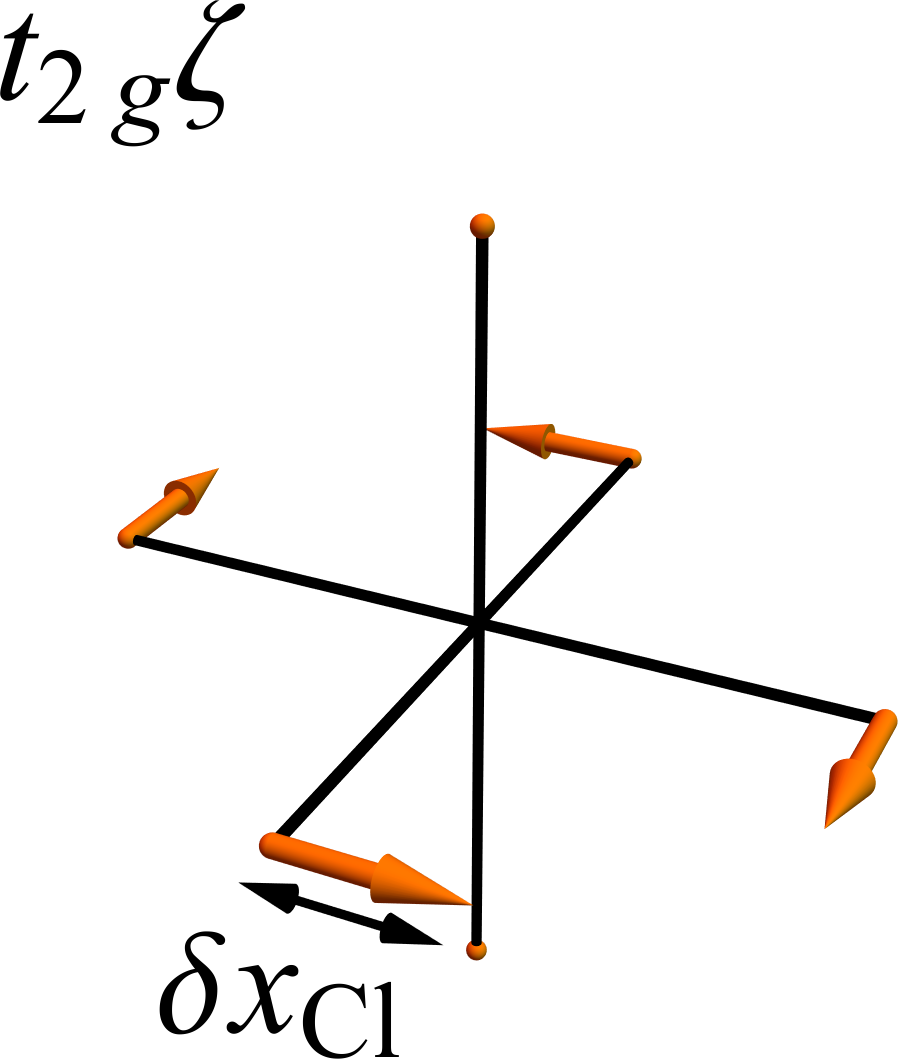}
&
~
&
\includegraphics[height=4.0cm]{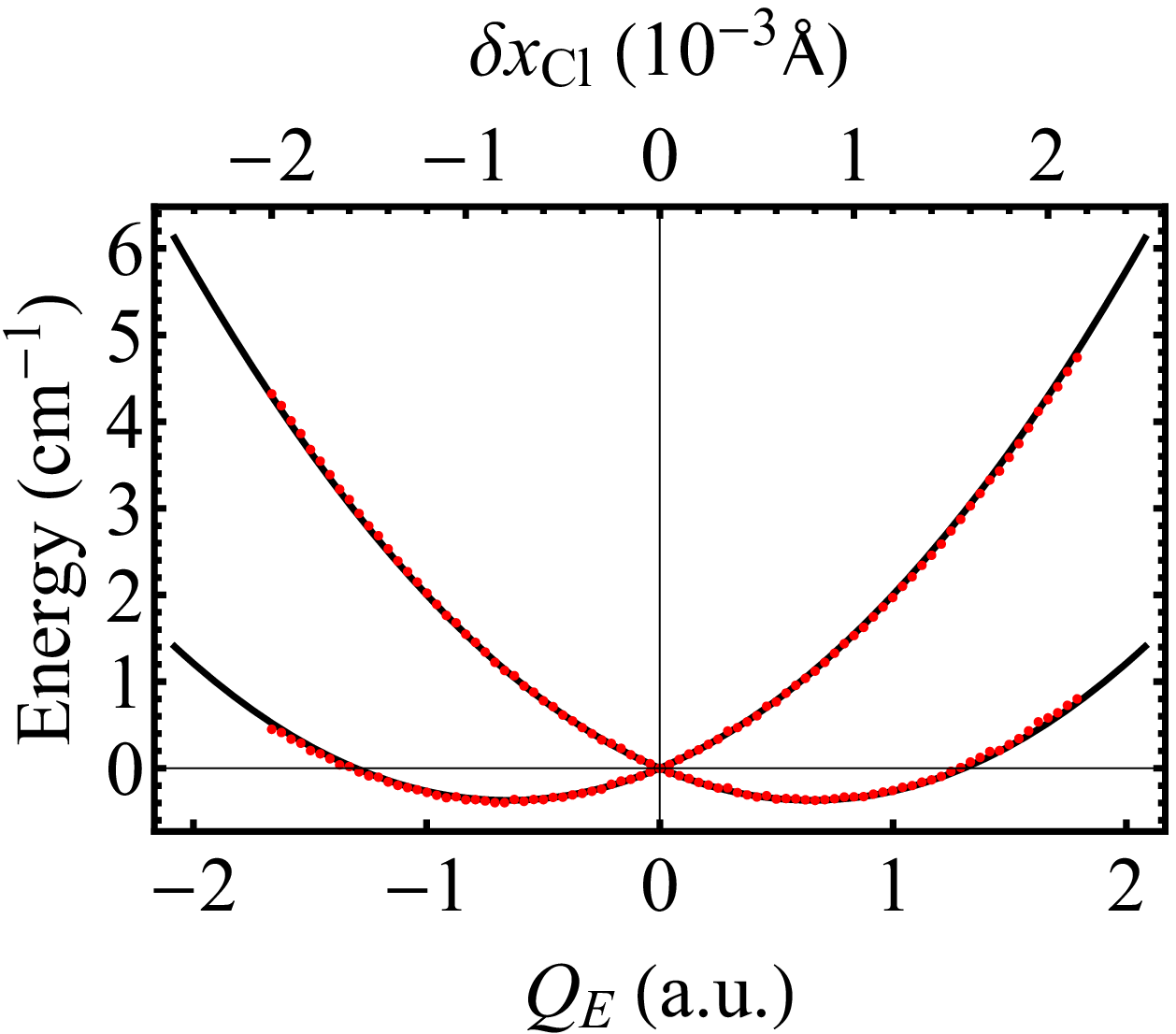}
&
~
&
\includegraphics[height=4.0cm]{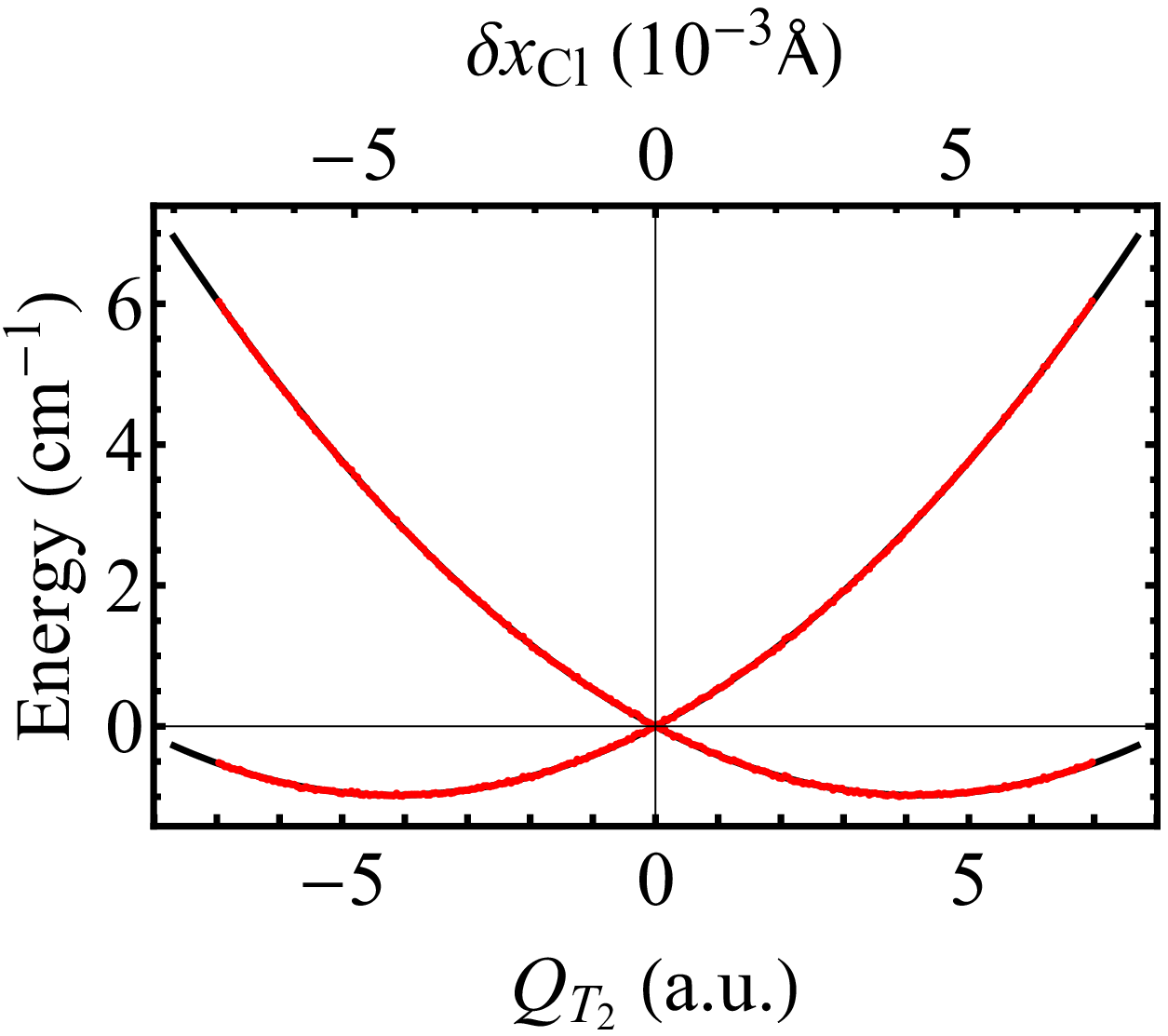}
\end{tabular}
\end{center}
\caption{
(color online)
(a-b) The $e_g\theta$ and $t_{2g}\xi$ vibrational modes. The $\Gamma_8$ APES of [NpCl$_6$]$^{2-}$ complex with respect to (c) $e_g\theta$ and (d) $t_{2g}\xi$ JT distortions (meV).
The red points are the {\it ab initio} results and the black solid lines are the APES fitted with Eq. (\ref{Eq:Uad}).
The JT distortions are measured by the displacement of Cl atom $\delta x_{\rm Cl}$ (10$^{-3}$ $\textrm{\AA}$) as shown in (a) and (b).
}
\label{FigK}
\end{figure*}

\begin{table}[tb]
\caption{{\it ab initio} (MB, DZP, TZP) and experimental (Exp.) $g$, $G$ and $g'$, $G'$ for Np$^{4+}$ ion.
MB, DZP, TZP indicate the basis set.}
\label{Table:gG}
\begin{ruledtabular}
\begin{tabular}{ccccc}
    &\multicolumn{3}{c}{{\it ab initio}} & Exp. \\
    & MB & DZP & TZP & \\
\hline
$g$ &   0.334  &   0.364  &   0.366  & \\
$G$ & $-1.860$ & $-1.918$ & $-1.925$ & \\
$g'$&   0.335  &   0.364  &   0.366  &   0.36 \\
$G'$& $-1.823$ & $-1.880$ & $-1.887$ & $-2.25$ \\
\end{tabular}
\end{ruledtabular}
\end{table}

\subsection{Definition of pseudospin from {\it ab initio} calculations}
\label{Sec:pseudospin}
The proper transformation of {\it ab initio} $\Gamma_8$ crystal-field states into $\Gamma_8$ pseudospin states is a crucial step for the unique definition of the pseudospin states. 
The crystal-field states are generally arbitrary linear combinations of the pseudospin states. 
However, in the present case, due to high symmetry of the complexes, the eigenstates of the magnetic moment projections along the main magnetic axes (tetragonal axes) correspond already to definite components of $\Gamma_8$ pseudospin states, $|\Phi_{\Gamma_8 M}\rangle$. 
This is also seen in the structure of component of the magnetic moments, Eqs. (\ref{Eq:mu_Gamma8}), (\ref{Eq:O}), which contains only powers of the corresponding pseudospin projection. 
We further denote the eigenstates of $\mu_z$ as $a$, $b$, $c$, $d$ in the increasing order of their eigenvalues. 
These eigenstates fulfill the time-reversal symmetry: under time-inversion, $a$ and $b$ transforms into $d$ and $c$, respectively.
However, there still remain eight possible assignments of the eigenstates to $|\Phi_{\Gamma_8 M}\rangle$. 
$(a,b,c,d)$ = $(\mp 3/2,\mp 1/2,\pm 1/2,\pm 3/2)$, $(\mp 3/2, \pm 1/2, \mp 1/2, \pm 3/2)$, $(\mp 1/2,\mp 3/2,\pm 3/2,\pm 1/2)$, and $(\mp 1/2, \pm 3/2, \mp 3/2, \pm 1/2)$. 
This issue is completely solved by analyzing the rotational symmetry properties of the states $a$-$d$.

For the assignment of multiconfigurational states $a$-$d$, it is sufficient to find symmetrized electron configurations which transform properly under symmetry operations. 
This is done straightforwardly by identifying the contributions of the true $S$ configurations in the eigenstates.
For example, in the case of $f^3$ system (Np$^{4+}$), the spin states $|SM\rangle$ of the admixed electronic term
${}^4A_2$ originating from $6p^3$ configuration of the actinide can easily be put in correspondence to $|\Gamma_8, M\rangle$
coincides with that of $\Gamma_8$ pseudospin states. 
Details of this assignment for Np$^{4+}$ and Ir$^{4+}$ are shown below.

\subsection{Weak vibronic coupling: Cs$_2$ZrCl$_6$:Np$^{4+}$}
The low-energy states of Np$^{4+}$ are described by crystal-field split $J=9/2$ atomic multiplet,
and it has been found that the ground state of the Np$^{4+}$ impurity is the $\Gamma_8$ state.
Our {\it ab initio} calculations reproduce the $\Gamma_8$ ground state.
The assignment of the $\Gamma_8$ state is done by comparing the ${}^4A_2$ configurations admixed to the $\Gamma_8$ states and
the relation between the $\Gamma_8$ representation and the product of the $A_2$ and the $\Gamma_8$ $(S=3/2)$ representations \cite{Koster1963}:
\begin{eqnarray*}
|\Gamma_8,\mp 3/2\rangle =  |{}^4A_2,\pm 1/2\rangle, \quad
|\Gamma_8,\mp 1/2\rangle = -|{}^4A_2,\pm 3/2\rangle, 
\end{eqnarray*}
where $|{}^4A_2,M\rangle$ is spin-orbital decoupled state and $M$ 
is the $z$-projection of spin. 
The $g$ and $G$ parameters calculated with three basis sets are shown in Table \ref{Table:gG}.
The results with the double zeta and triple zeta basis sets are close to each other.
Thus, hereafter, we use the double zeta basis set for the calculations of Np$^{4+}$.

The vibronic coupling constants $k_\Gamma$ and the frequencies of NpCl$_6^{2-}$
are derived by fitting the model APES, Eq. (\ref{Eq:Uad}), to the APES obtained in {\it ab initio} calculations.
Since the octahedral system has only one $e_g$ and one $t_{2g}$ vibrational modes, the corresponding displacements are determined by symmetry
(Fig. \ref{FigK} (a) and (b)) \cite{Bersuker1989, Bersuker2006}.
The APES with respect to the $e_g \theta$ and the $t_{2g} \zeta$ JT distortions are shown in Fig. \ref{FigK} (c) and (d), respectively
\footnote{
The atomic unit of mass-weighted coordinate is $\sqrt{m_e} a_0$, where $m_e$ is electron mass, and $a_0$ is Bohr radius.
}.
The red points are the {\it ab initio} APES with respect to the
JT distortion and the black solid lines are the simulations with Eq. (\ref{Eq:Uad}).
The best fitted frequencies and the dimensionless vibronic coupling constants are $\omega_{E}=618$ cm$^{-1}$, $\omega_T=157$ cm$^{-1}$, $k_E=0.035$, and $k_T=0.111$.
Although the existence of the JT effect in this complex has been anticipated \cite{Bray1978, Warren1983}, its role is found now negligible. 
One of the reason why the vibronic coupling is so weak is explained by the fact that 
the ground molecular term in the absence of spin-orbit coupling is the orbitally non-degenerate $^4A_2$ according to CASSCF calculations. 
The second reason is the relatively localized nature of the $5f$ orbitals.

With the use of the vibronic coupling constants and vibrational frequencies, 
we calculated the vibronic states by numerical diagonalization of the JT Hamiltonian (see Sec. \ref{Sec:Vibronic}).
The vibronic reduction factors are $K_{11} = 0.985$, $K_{12} = -0.007$, $K_{21} = -0.003$ and $K_{22} = 0.979$.
The obtained $g'$ and $G'$ (\ref{Eq:gG_DJT}) are given in Table \ref{Table:gG}.
The calculated $g'$ and $G'$ are close to $g$ and $G$, respectively, as a result of weak vibronic coupling in the $\Gamma_8$ multiplet. 
They are in good agreement with the experimental values listed in Table \ref{Table:gG} 
\footnote{
In Ref. \cite{Bernstein1979}, the eigenstates of $\mu_z$ with large and small eigenvalues are assigned to 
\unexpanded{$|\tilde{S} \pm 3/2\rangle$} and \unexpanded{$|\tilde{S} \pm 1/2\rangle$}, which is different from our definition using group theory. 
Here, their $g$ and $G$ are recalculated based on our definition.
}.
The obtained discrepancy for $G$ ($G'$) may be attributed to the insufficient accuracy of {\it ab initio} calculations, which did not include the CASPT2 step. 

A similar situation takes place in [(C${\mathrm{H}}_{3}$)$_{4}\mathrm{N}$]$_{2}\mathrm{Np}$${\mathrm{Cl}}_{6}$ \cite{Karraker1980} and in NpO$_2$.
Although in the latter system the JT effect is stronger than in NpCl$_6^{2-}$ due to the presence of more covalent oxygen ligands, the ratio between $E_\text{JT}$ and vibrational quantum corresponding to the JT-active distortions will be still very small 
as in isostructural UO$_2$ \cite{Mironov2003}.
On this reason, according to Eqs. (\ref{Eq:K_weak}), the manifestation of JT effect on the magnetic moments is expected to be unimportant.

\subsection{Strong vibronic coupling: ThO$_2$:Ir$^{4+}$}

\begin{figure*}[tb]
\begin{center}
\begin{tabular}{ccccccc}
\multicolumn{1}{l}{(a)} & ~ &
\multicolumn{1}{l}{(b)} & ~ &
\multicolumn{1}{l}{(c)} & ~ &
\multicolumn{1}{l}{(d)} \\
\includegraphics[height=3.4cm]{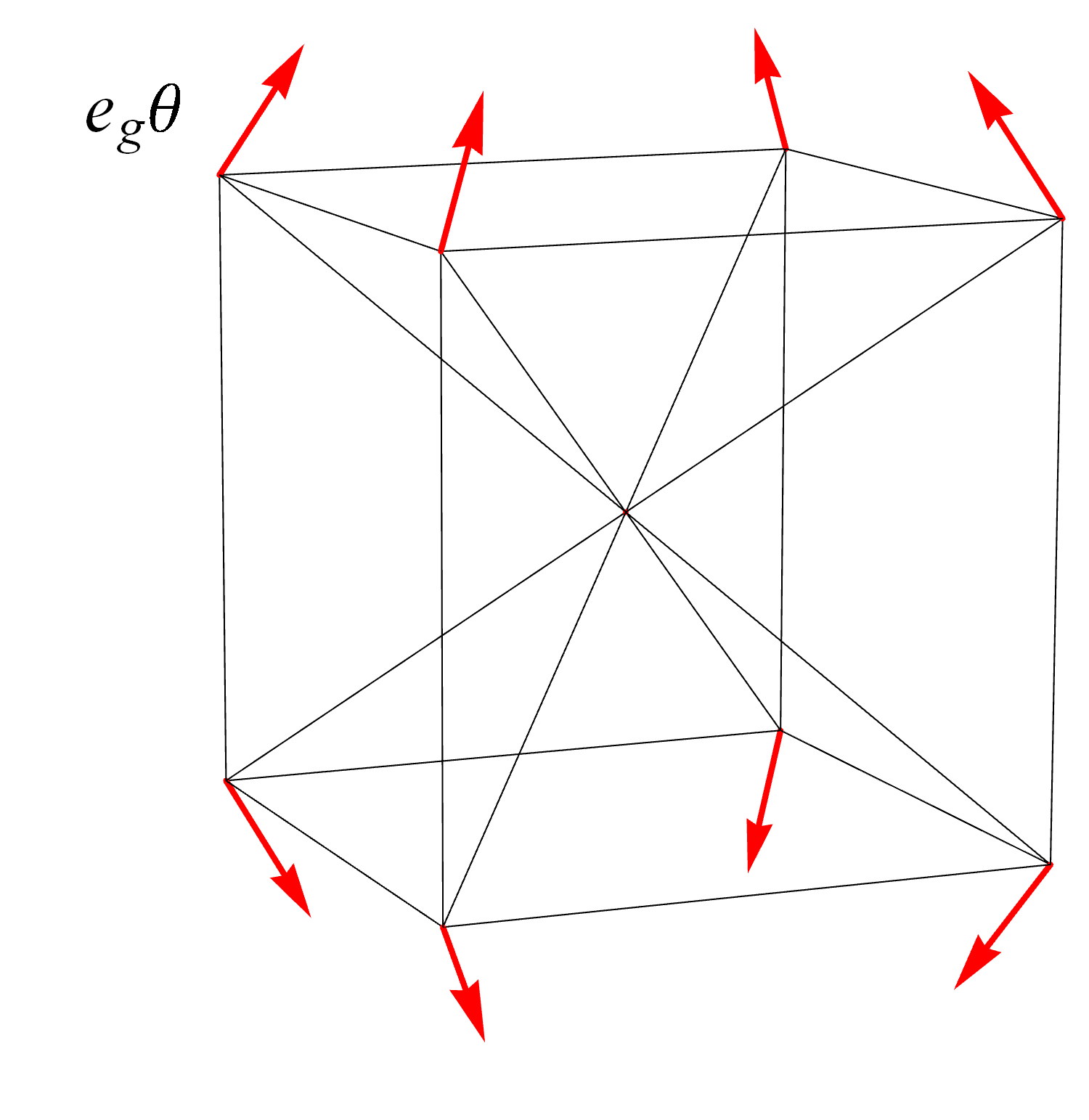}
&
~
&
\includegraphics[height=3.4cm]{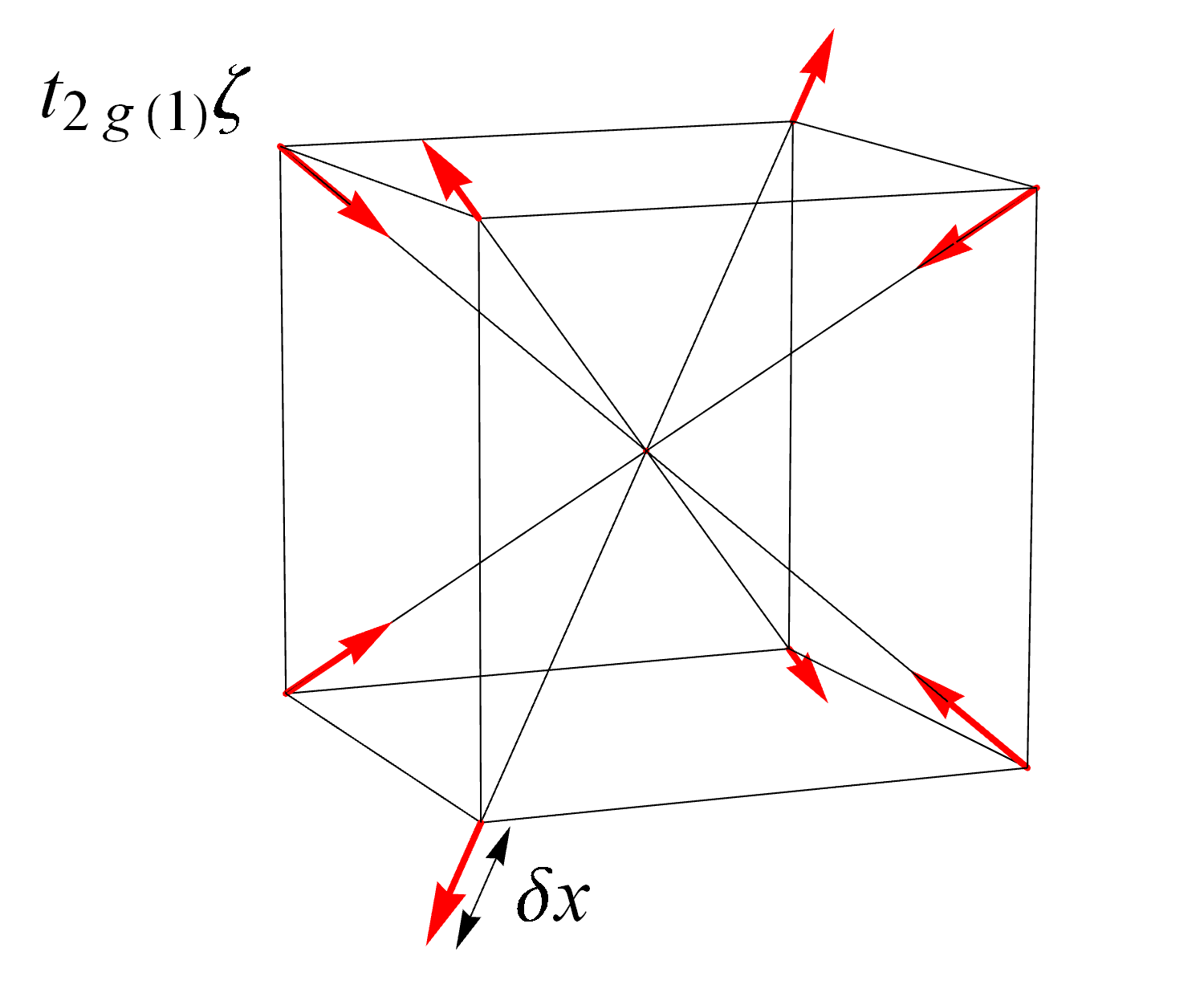}
&
~
&
\includegraphics[height=3.4cm]{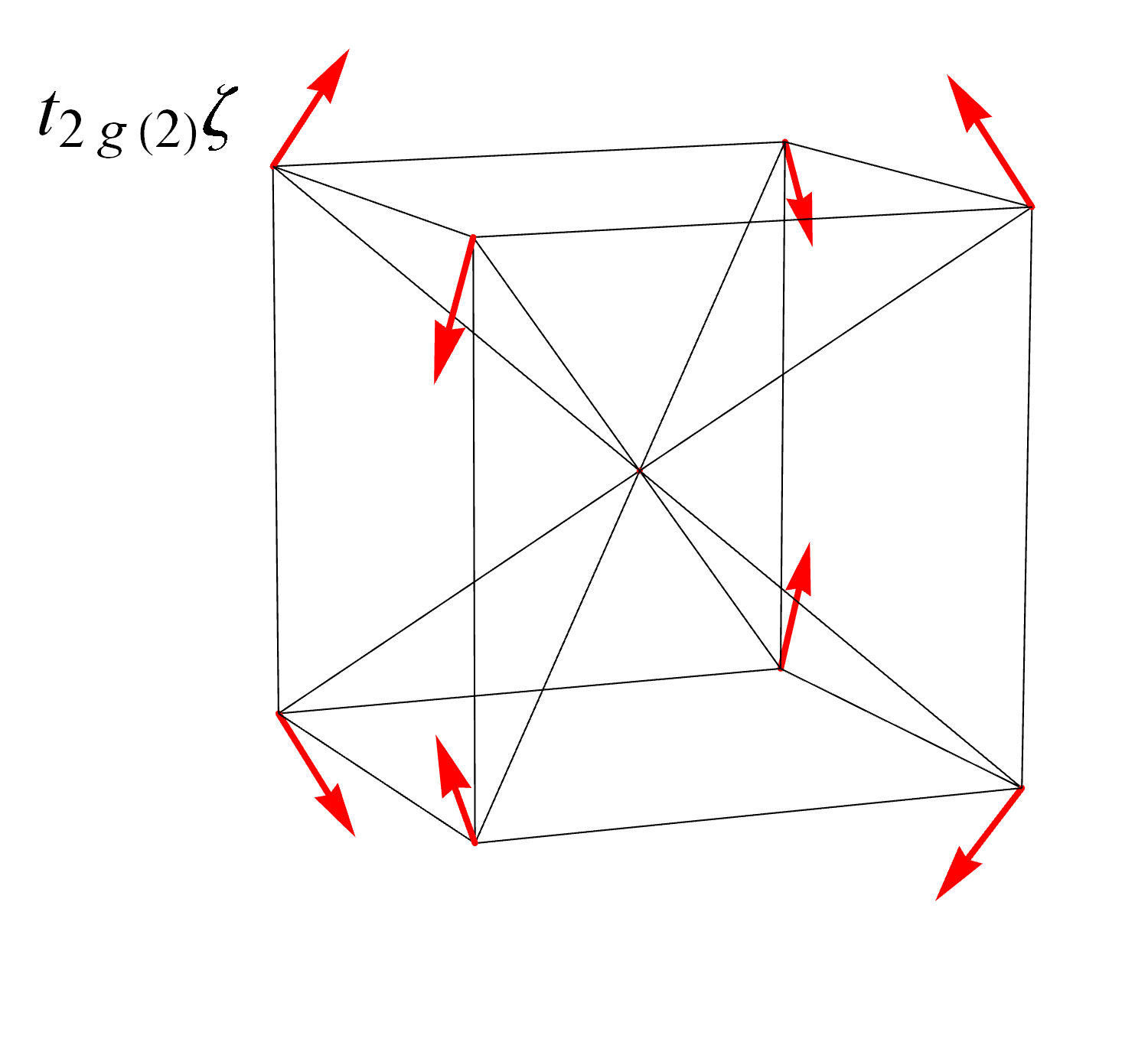}
&
~
&
\includegraphics[height=4.2cm]{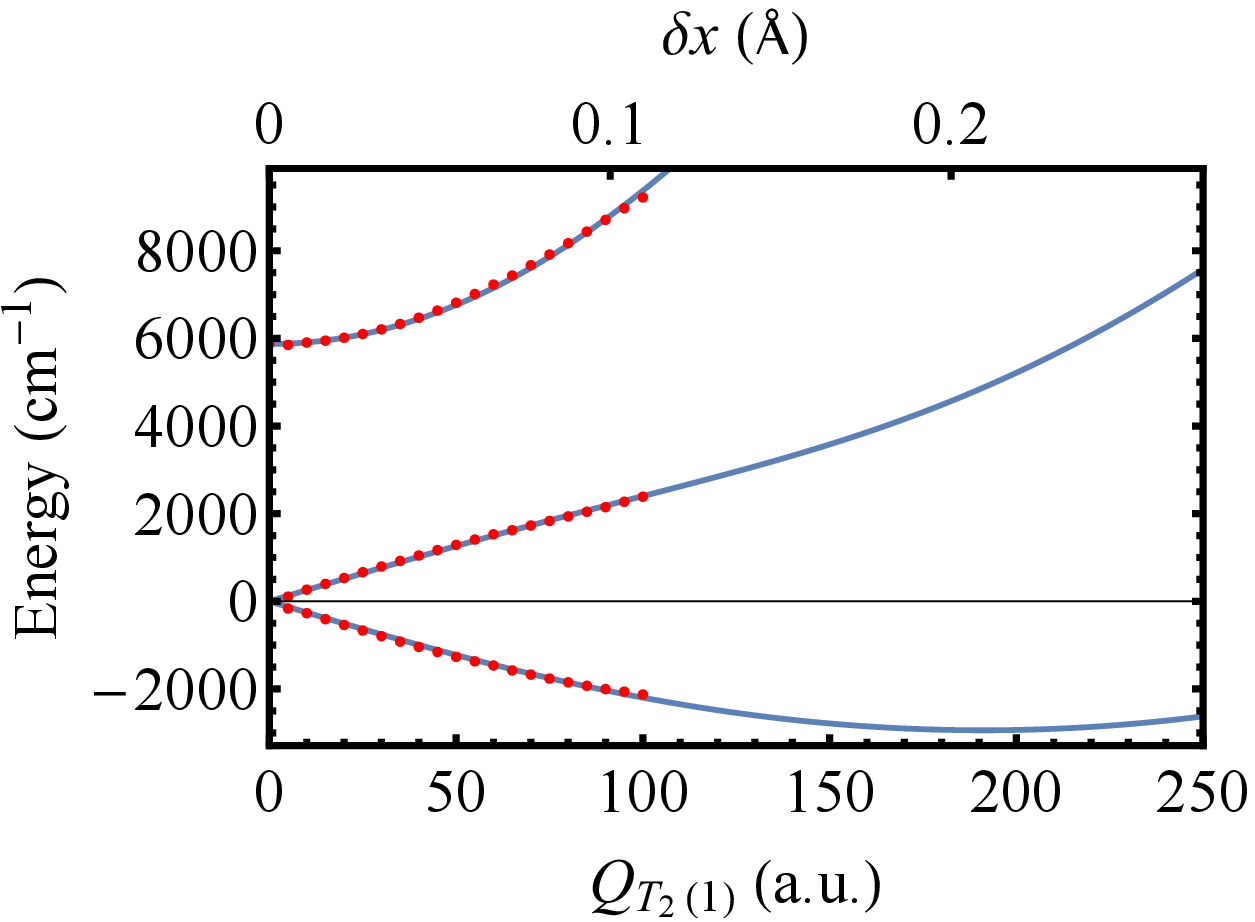}
\\
\end{tabular}
\end{center}
\caption{(color online)
(a-c) The $e_{g}\theta$, $t_{2g}(1)\zeta$ and $t_{2g}(2)\zeta$ distortions of a cube.
(d) The APES with respect to $t_{2g(1)}\zeta$ JT distortion. The red points correspond to the {\it ab initio} calculations, while the black lines are the fitted APES. $\delta x$ shows the displacement of O atoms in (b).
}
\label{Cub_dist}
\end{figure*}

Stronger vibronic coupling is expected when the overlap of the magnetic orbitals and the ligand orbitals is large.
This seems to be the case of $5d$ ions such as Ir$^{4+}$. 
In a cubic environment, the atomic $d$ shell splits into $e_g$ and $t_{2g}$ levels, the latter being higher in energy.
Hence, in the case of $5d^5$ configuration of Ir$^{4+}$ the $e_g$ levels are fully occupied and the $t_{2g}$ levels are singly occupied.
The CASSCF/CASPT2 calculations confirm this picture, showing that the ground state of Ir$^{4+}$ center corresponds to the $S=1/2$ (without spin-orbit coupling), 
mainly coming from the configuration $e_g^4t_{2g}^1$.
In the presence of spin-orbit coupling, the ${}^2T_{2g}$ term splits into ground $\Gamma_8$ and excited $\Gamma_7$ states. 
The ground $\Gamma_8$ multiplet states are assigned by analyzing the contributions of the ${}^2T_{2g}$ 
in comparison with the symmetry relation between the direct product $T_{2g} \otimes \Gamma_6$ ($S=1/2$) and the $\Gamma_8$ representation \cite{Koster1963}: 
\begin{eqnarray*}
|\Gamma_8,\mp 3/2\rangle &=& \mp \frac{i}{\sqrt{6}} |{}^2T_{2g}\xi, \mp 1/2\rangle 
                               + \frac{1}{\sqrt{6}} |{}^2T_{2g}\eta, \mp 1/2\rangle 
\nonumber\\
                            &+& i\sqrt{\frac{2}{3}} |{}^2T_{2g}\zeta, \pm 1/2\rangle,
\nonumber\\
|\Gamma_8,\mp 1/2\rangle &=& \pm \frac{i}{\sqrt{2}} |{}^2T_{2g}\xi, \pm 1/2 \rangle
                               - \frac{1}{\sqrt{2}} |{}^2T_{2g}\eta, \pm 1/2\rangle,
\end{eqnarray*}
where $|{}^2T_{2g} \gamma,M\rangle$ is the term's wave function, and $\gamma$ ($= \xi, \eta, \zeta$) and $M$ ($= \pm 1/2$) are the orbital and spin components, respectively.
The $g$ and $G$ parameters are obtained as $-0.076$ and 0.169, respectively.
The small $g$ and $G$ are explained by the partial cancellation of the orbital and spin angular momenta \cite{Kotani1949}.

The cubic system contains one $e_g$ (Fig. \ref{Cub_dist} (a)) and two $t_{2g}$ (Fig. \ref{Cub_dist} (b), (c)) vibrational modes.
The APES with respect to the $t_{2g}(1)$ distortion shows a very large JT stabilization energy, Fig. \ref{Cub_dist} (d),
whereas those for the $e_g$ and the $t_{2g}(2)$ are negligibly small (APES are not shown).
The red points in Fig. \ref{Cub_dist} (d) correspond to the {\it ab initio} APES, which shows huge Jahn-Teller splitting.
The strong vibronic coupling to the $t_{2g}(1)$ mode is due to both the delocalization of the $5d$ orbitals and the distortion along the Ir-O bond (see Fig. \ref{Cub_dist} (b)), while the couplings to the $e_g$ and the $t_{2g}(2)$ modes are small because the displacements are perpendicular to the Ir-O bonds (Fig. \ref{Cub_dist} (a), (c)).
Figure \ref{Cub_dist}(d) also shows that the upper level of the $\Gamma_8$ state does not follow a parabola as expected from Eq. (\ref{Eq:Uad}).
The discrepancy is expected to originate from the vibronic coupling between the ground $\Gamma_8$ states and the excited $\Gamma_7$ states.
Accordingly, we include in our analysis of the APES the excited Kramers doublet. 
The JT Hamiltonian in this case consists of the $\Gamma_8$ part given in Eq. (\ref{Eq:HJT}),
the vibrational Hamiltonian ($\Delta E+\sum_{\Gamma \gamma} \frac{1}{2}{\omega'_\Gamma}^2Q_\gamma^2$) for the $\Gamma_7$ multiplets
($\Delta E$ is the excitation energy of the $\Gamma_7$ level with respect to the ground $\Gamma_8$ level, $\omega'_\Gamma$ is the frequency),
and the cross vibronic terms between the $\Gamma_8$ and the $\Gamma_7$ multiplets:
\begin{eqnarray}
 U' &=& \sum_{\Gamma \gamma} \left( V'_\Gamma C'_{\Gamma \gamma} Q_{\Gamma \gamma} + \text{H.c.}\right),
\end{eqnarray}
where $V'_{\Gamma}$ are real off-diagonal vibronic coupling constants, and $C'_{\Gamma \gamma}$ are defined by
\begin{eqnarray}
 C'_{E\theta} &=&
 \begin{pmatrix}
  0 & -1 \\
  0 &  0 \\
  0 &  0 \\
  1 &  0
 \end{pmatrix},
\quad
 C'_{E\epsilon} =
 \begin{pmatrix}
  0 & 0 \\
  1 & 0 \\
  0 & -1 \\
  0 & 0
 \end{pmatrix},
\nonumber\\
 C'_{T_2\xi} &=&
 \begin{pmatrix}
  i\frac{\sqrt{3}}{2} & 0 \\
  0 & \frac{i}{2} \\
  -\frac{i}{2} & 0 \\
  0 & -i\frac{\sqrt{3}}{2}
 \end{pmatrix},
\quad
 C'_{T_2\eta} =
 \begin{pmatrix}
  \frac{\sqrt{3}}{2} & 0 \\
  0 & \frac{1}{2}  \\
  \frac{1}{2} & 0  \\
  0 & \frac{\sqrt{3}}{2}
 \end{pmatrix},
\nonumber\\
 C'_{T_2\zeta} &=&
 \begin{pmatrix}
   0 & 0 \\
   i & 0 \\
   0 & i \\
   0 & 0
 \end{pmatrix}.
\label{Eq_Cgamma7}
\end{eqnarray}
The basis for the rows and the columns of $C'_{\Gamma\gamma}$ are the $\Gamma_8$ and $\Gamma_7$ states in the increasing order of $M$.
Applying the second order perturbation theory, the APES are expressed as 
\begin{eqnarray}
 U^{\Gamma_8}_{\pm}(Q) &=& \frac{1}{2}\omega_T^2Q_{T_{2g}\zeta}^2 \pm V_{T_2} Q_{T_{2g}\zeta}
\nonumber\\
 &+& \frac{\frac{1}{2}{V^{\prime 2}_{T_2}} Q_{T_2\zeta}^2}
   {-\Delta E + \frac{1}{2}(\omega_{T_2}^2 - \omega^{\prime 2}_{T_2}) Q_{T_2\zeta}^2 \pm V_{T_2} Q_{T_2\zeta}},
\nonumber\\
U^{\Gamma_7}(Q)&=& \Delta E + \frac{1}{2}{\omega^{\prime 2}_{T_2}} Q_{T_2\zeta}^2
\nonumber\\
&+& \sum_{\sigma = \pm} \frac{\frac{1}{2}{V^{\prime 2}_{T_2\zeta}} Q_{T_2\zeta}^2}
  {\Delta E - \frac{1}{2} (\omega^2_{T_2} - {\omega^{\prime 2}_{T_2}}) Q_{T_2\zeta}^2 + \sigma V_{T_2} Q_{T_2\zeta}}.
\nonumber\\
 \label{Eq_U_Gamma78}
\end{eqnarray}
The best fitting of the {\it ab initio} APES to Eq. (\ref{Eq_U_Gamma78}) 
is given by $V_{T_2}=1.17\times 10^{-4}$ a.u., $\omega_{T_2}=191.32$ cm$^{-1}$, $V'_{T_2}=1.37\times 10^{-4}$ a.u. and $\omega'_{T_2}=298.63$ cm$^{-1}$ 
(solid lines in Fig. \ref{Cub_dist}(d)).
The calculated dimensionless vibronic coupling constant $k_{T_2}=4.54$ indicates that Ir$^{4+}$ will show strong JT effect
($E_{T_2(1)}$ exceeds $\hslash \omega_{T_2(1)}$ by several times).

With the obtained vibronic coupling constants, we calculated the reduction factors as follows, $K_{11}=0.60$, $K_{12}=-0.079$, $K_{21}=-0.22$ and $K_{22}=0.40$.
For simplicity, the $\Gamma_7$ states were not taken into account in the simulation.
Using the reduction factors $K_{ij}$ we obtain $g' = -0.083$ and $G' = 0.074$. 
As expected for a large vibronic coupling, $G'$ differs much from $G$ ($g = -0.076$ and $G =$ 0.169).

\section{Conclusions}
In this work, we investigate thoroughly the interplay between the Zeeman interaction and the Jahn-Teller effect in the $\Gamma_8$ multiplet.
Combining the theory and {\it ab initio} quantum chemistry calculations, we have demonstrated the role of the Jahn-Teller effect on the magnetic moment.
This is achieved by using an {\it ab initio} methodology of the derivation of Zeeman pseudospin Hamiltonian and of the vibronic parameters of the $\Gamma_8 \otimes (e \oplus t_2)$ JT Hamiltonian. 
The main conclusions are the following:
\begin{enumerate}
 \item Dynamical JT effect can modify not only the absolute values of the parameters of Zeeman pseudospin Hamiltonian
       but also their signs with respect to pure electronic case.
 \item In the presence of the static JT distortion, the nature of the magnetic moments depends on the type of distortion as well as on the values of the $g$ factors.
       In particular, the sign of $g_Xg_Yg_Z$ can change due to the rotation of the JT distortion, which could be experimentally observed by using circular polarized magnetic field or in hyperfine spectra.
\end{enumerate}
The strong Jahn-Teller coupling of Ir$^{4+}$ impurity in a cubic site predicted by our {\it ab initio} calculations is intriguing because of the lack of the experimental study of the strong Jahn-Teller effect in $\Gamma_8$ systems.
The present {\it ab initio} approach 
allowing for quantitative study of the interplay of the Zeeman and vibronic interactions will be useful for the study of various $\Gamma_8$ 
and other multiplet states in complexes and correlated materials. 
Further applications of the present methodology will be discussed elsewhere.

\section*{Acknowledgment}
N. I. is Japan Society for the Promotion of Science (JSPS) Overseas Research Fellow.
He would like also to acknowledge the financial support from the Fonds Wetenschappelijk Onderzoek - Vlaanderen (FWO) and the GOA grant from KU Leuven.
V. V. and L. U. are postdoctoral fellows of FWO.

\appendix

\section{Vibronic state in the weak vibronic coupling}
\label{Sec:Vibronic_pert}
Within the first order of perturbation theory,
the ground vibronic states are given as
\begin{eqnarray}
 |\Psi_{\Gamma_8 \frac{3}{2}}\rangle
 &=& N \left[ |\Phi_{\Gamma_8\frac{3}{2}}\rangle|00000\rangle
 - \frac{k_E}{\sqrt{2}} |\Phi_{\Gamma_8\frac{3}{2}}\rangle|10000\rangle
 \right.
\nonumber\\
 &-& \frac{k_E}{\sqrt{2}} |\Phi_{\Gamma_8-\frac{1}{2}}\rangle|01000\rangle
 + \frac{ik_{T_2}}{\sqrt{2}} |\Phi_{\Gamma_8\frac{1}{2}}\rangle |00100\rangle
\nonumber\\
 &+&
 \left.
   \frac{k_{T_2}}{\sqrt{2}} |\Phi_{\Gamma_8\frac{1}{2}}\rangle |00010\rangle
 + \frac{ik_{T_2}}{\sqrt{2}} |\Phi_{\Gamma_8-\frac{1}{2}}\rangle|00001\rangle
 \right],
\nonumber\\
 |\Psi_{\Gamma_8 \frac{1}{2}}\rangle
 &=& N\left[|\Phi_{\Gamma_8\frac{1}{2}}\rangle|00000\rangle
 + \frac{k_E}{\sqrt{2}} |\Phi_{\Gamma_8\frac{1}{2}}\rangle|10000\rangle
 \right.
\nonumber\\
 &-& \frac{k_E}{\sqrt{2}} |\Phi_{\Gamma_8-\frac{3}{2}}\rangle|01000\rangle
 + \frac{ik_{T_2}}{\sqrt{2}} |\Phi_{\Gamma_8\frac{3}{2}}\rangle |00100\rangle
\nonumber\\
 &-&
 \left.
   \frac{k_{T_2}}{\sqrt{2}} |\Phi_{\Gamma_8\frac{3}{2}}\rangle |00010\rangle
 - \frac{ik_{T_2}}{\sqrt{2}} |\Phi_{\Gamma_8-\frac{3}{2}}\rangle|00001\rangle
 \right],
\nonumber\\
\label{Eq:vibronic_weak}
\end{eqnarray}
where $N$ is the normalization constant, $N = 1/\sqrt{1+k_E^2+3k_{T_2}^2/2}$.
Substituting Eq. (\ref{Eq:vibronic_weak}) into 
Eq. (\ref{Eq:Kij}) and solving that system of equation, we obtain Eq. (\ref{Eq:K_weak}). 


%

\end{document}